\documentclass[aps,twocolumn,reprint,superscriptaddress]{revtex4-2}
\usepackage{amsmath}
\usepackage{amssymb}
\usepackage[latin9]{inputenc}
\usepackage{graphicx}
\usepackage{hyperref}
\usepackage{physics}
\usepackage{changepage}
\usepackage[color= blue!20]{todonotes}
\usepackage[normalem]{ulem}
\usepackage{color}
\usepackage{ulem}
\usepackage[symbol]{footmisc}
\usepackage{xcolor}

\definecolor{myBlue}{rgb}{0.29,0.47,1}

\hypersetup{
	colorlinks,
	linkcolor={blue!70!black},
	citecolor={blue!70!black},
	urlcolor={blue!70!black}
}

\begin{document}

\title{Photon-Pressure with an Effective Negative Mass Microwave Mode}

\author{I.~C.~Rodrigues}\email{icorveira@phys.ethz.ch}
\affiliation{Kavli Institute of Nanoscience, Delft University of Technology, PO Box 5046, 2600 GA Delft, The Netherlands}
\affiliation{Department of Physics, ETH Zürich, Zürich, Switzerland}
\author{G.~A.~Steele}\email{g.a.steele@tudelft.nl}
\affiliation{Kavli Institute of Nanoscience, Delft University of Technology, PO Box 5046, 2600 GA Delft, The Netherlands}
\author{D.~Bothner}\email{daniel.bothner@uni-tuebingen.de}
\affiliation{Physikalisches Institut, Center for Quantum Science (CQ) and LISA$^+$, Universit\"at T\"ubingen, 72076 T\"ubingen, Germany}
\affiliation{Kavli Institute of Nanoscience, Delft University of Technology, PO Box 5046, 2600 GA Delft, The Netherlands}

\begin{abstract}
	Harmonic oscillators belong to the most fundamental concepts in physics and are central to many current research fields such as circuit QED, cavity optomechanics and photon-pressure systems.
	Here, we engineer a microwave mode in a superconducting LC circuit that mimics the dynamics of a negative mass oscillator, and couple it via photon-pressure to a second low-frequency circuit.
	We demonstrate that the effective negative mass dynamics lead to an inversion of dynamical backaction and to sideband-cooling of the low-frequency circuit by a blue-detuned pump field, which can be intuitively understood by the inverted energy ladder of a negative mass oscillator.

\end{abstract}

\maketitle

The harmonic oscillator (HO) is one of the most fundamental models in physics and can be used to describe many kinds of systems, most prominently mechanical oscillators and electrical resonant circuits, but also optical cavities, acoustic crystal vibrations or collective spin oscillations in magnets.
HOs also play a crucial role for the development of quantum technologies, of which some of the most relevant are circuit quantum electrodynamics (cQED) \cite{You11, Blais21} and cavity optomechanics \cite{Aspelmeyer14, Barzanjeh22} as well as its cQED equivalent, photon-pressure systems \cite{Johansson14, Eichler18, Bothner21}.
In nearly all experimental cases HOs have a positive ``mass" (positive capacitance in LC circuits).
Note that we use the term ``mass" in a generalized sense, i.e., for any quantity that describes the inertia of the HO.
However, in addition to theoretical considerations \cite{Glauber86, Hammerer09, Tsang10, Tsang12, Zhang13, Motazedifard16} there have been experimental reports of effective negative mass HOs realized through spin ensembles \cite{Julsgaard01, Wasilewski10, Moller17, Kohler18}, through the common modes of two micromechanical oscillators \cite{deLepinay21}, and in multimode electromechanical systems \cite{Bernier18}.
In these works, the negative mass has fascinating practical consequences such as providing quantum-mechanics-free or backaction-free subspaces \cite{Tsang12}  and enabling entanglement between distinct oscillators \cite{Julsgaard01, deLepinay21}.
Despite these promising perspectives, it is very challenging to experimentally realize negative effective mass oscillators and new approaches are under investigation \cite{Junker22}.
Here, we present a simple method to create an effective negative mass HO which only requires a Kerr nonlinearity and is therefore not limited to specific platforms.
The effective negative mass HO is prepared by strongly driving a weakly nonlinear superconducting LC circuit, which leads to a susceptibility inversion compared to the positive mass case.
Similar driving schemes in nonlinear systems have lately been implemented with both LC circuits and mechanical oscillators \cite{Huber20, FaniSani21}, but it has not been demonstrated yet that such an approach creates a dynamically stabilized analog of a negative mass mode.
The integration of this mode into a photon-pressure device allows us to use the interaction with a low-frequency (LF) circuit as a probe for the effective mode mass through dynamical backaction.
Most strikingly, we find that dynamical backaction effects get inverted compared to positive mass modes, which leads to sideband-cooling of the LF circuit by a blue-detuned pump field.

\begin{figure*}
	\centerline{\includegraphics[trim = {0.5cm, 0.5cm, 1.0cm, 0cm}, clip=True,width=\textwidth]{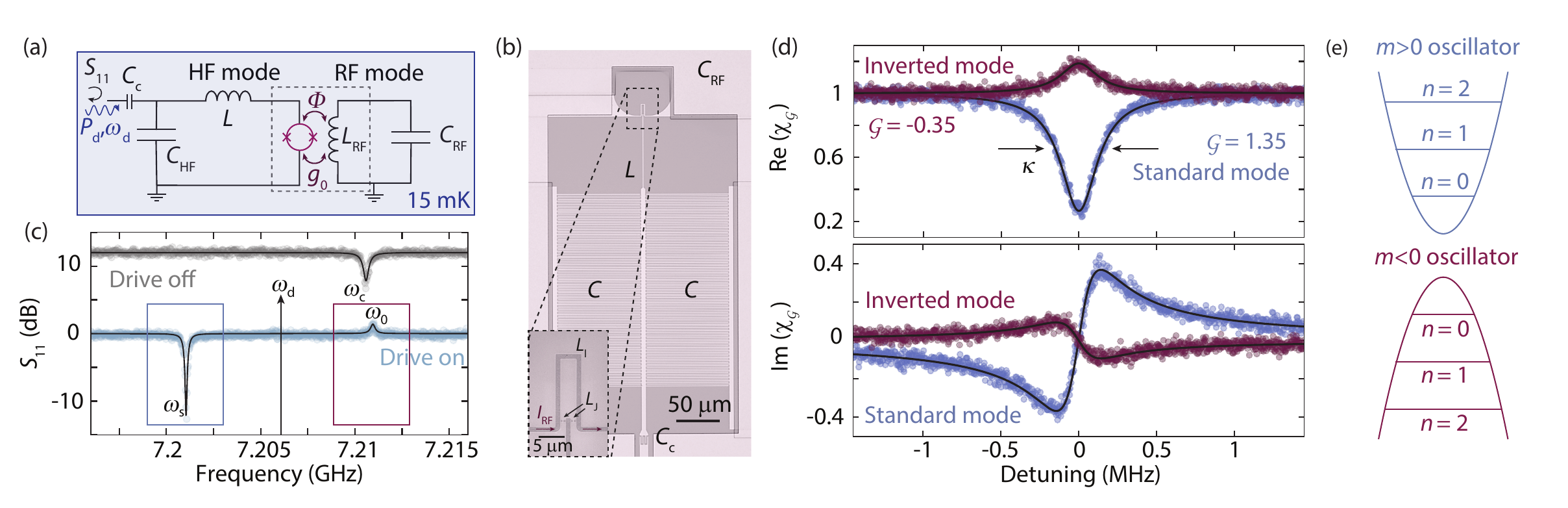}}
	\caption{\textsf{\textbf{Engineering an effective negative mass microwave mode in a strongly driven photon-pressure circuit.} (a) Circuit schematic.  Linear inductances and capacitances are $L, C_\mathrm{HF} = 2C, L_\mathrm{RF}, C_\mathrm{RF}$, the single-photon coupling rate is $g_0$ and the HF mode is coupled to a feedline by means of a coupling capacitance $C_\mathrm{c}$. The HF circuit is driven with a strong, near-resonant tone at $\omega_\mathrm{d}$ and with power $P_\mathrm{d}$ and simultaneously the reflection response $S_{11}(\omega)$ is tracked with a small probe signal. (b) Images of the device. Bright parts are aluminum, dark and transparent parts are silicon. Inset shows scanning electron microscopy (SEM) image of the coupling region. Junctions are labelled with $L_\mathrm{J}$, loop inductance by $L_\mathrm{l}$ and the RF mode current is indicated by $I_\mathrm{RF}$. In the SEM image darker parts are Al, brighter parts Si. (c) Reflection $S_{11}$ both without and with strong drive in direct comparison, lines are fits. Undriven curve is offset by $+12\,$dB. The undriven resonance frequency is $\omega_\mathrm{c}$, the driven response shows two modes at $\omega_\mathrm{s}$ and $\omega_0$. (d) Real and imaginary parts of the mode susceptibilities $\chi_\mathcal{G}$ of the two modes at $\omega_\mathrm{s}$ and $\omega_0$ vs their corresponding detunings $\varDelta_0, \varDelta_\mathrm{s}$. Lines are fits with Eq.~(\ref{eqn:Gsus}). (e) Potential and energy levels of a positive mass ($\mathcal{G}>0$, top) and a negative mass ($\mathcal{G}<0$, bottom) HO.}}
	\label{fig:Fig1}
\end{figure*}

To explore the (classical) phenomenology of a negative mass HO, we consider the susceptibility $\chi_m(\omega)$, which describes the response of a low-loss mechanical oscillator with mass $m$ to an external excitation $F_\mathrm{ex}(\omega)$ in frequency space
\begin{equation}
	\chi_m(\omega) = \frac{1}{2 m \omega_0}\frac{1}{\frac{\kappa}{2} + i(\omega - \omega_0)},
	\label{eqn:first}
\end{equation} 
i.e., $x(\omega) = -i\chi_m(\omega) F_\mathrm{ex}(\omega)$.
Here, $\omega$ is the excitation frequency, $\omega_0$ is the resonance frequency, and $\kappa$ is the oscillator decay rate.
With a negative mass, however, one obtains a HO with a susceptibility $\chi_-(\omega) = -\chi_+(\omega)$, where $\chi_+(\omega)$ is the susceptibility of a positive mass oscillator \cite{Supplement}.
At first glance, this may not look particularly striking, as it introduces only a phase shift of $\pi$ compared to a positive mass oscillator, but it is the essence of an effective negative mass \cite{Supplement}.
Below, we will demonstrate that integrating such an inverted susceptibility oscillator and its corresponding phase-shifted response into a system of photon-pressure circuits leads to dramatic consequences in the interaction of the two circuits.
Before we dive into the details of our experiment, we  introduce a suceptibility expression for a more general HO
\begin{equation}
	\chi_\mathcal{G}(\omega) = \frac{\mathcal{G}}{\frac{\kappa}{2} + i\left(\omega - \omega_0\right)}.
	\label{eqn:Gsus}
\end{equation}
Here, we have chosen a convention without the prefactor $1/(2m\omega_0)$ of Eq.~(\ref{eqn:first}), but added a dimensionless parameter $\mathcal{G}$ in the numerator, which becomes negative for a negative mass oscillator.
The case $\mathcal{G} > 1$ represents intracavity amplification, recently considered in Ref.~\cite{Rodrigues22}, and here we explore the implications of $\mathcal{G} < 0$.
Our device combines a superconducting radio-frequency (RF) circuit with a high-frequency (HF) quantum interference circuit, which are coupled to each other via a magnetic flux-tunable photon-pressure interaction, cf. Fig.~\ref{fig:Fig1}(a), (b) and Refs.~\cite{Rodrigues22, Rodrigues21}.
The RF circuit has a resonance frequency $\varOmega_0 = 2\pi \cdot 452\,$MHz and a linewidth $\varGamma_0 = 2\pi\cdot 45\,$kHz.
The undriven HF cavity has a resonance frequency of $\omega_\mathrm{c} = 2\pi \cdot 7.211\,$GHz, a total (external) linewidth of $\kappa = 2\pi\cdot 420\,$kHz ($\kappa_\mathrm{e} = 2\pi\cdot 78\,$kHz).
We flux-bias the device at an operation point with a single-photon coupling rate $g_0 = 2\pi\cdot 175\,$kHz and an HF cavity Kerr constant $\mathcal{K} = -2\pi \cdot 6.6\,$kHz, which originates from the integrated constriction-type Josephson junctions.
All experiments have been conducted in a dilution refrigerator at a base temperature of $T_\mathrm{b} \approx 15\,$mK.
More details can be found in Refs.~\cite{Rodrigues22, Rodrigues21}.

\begin{figure*}
	\centerline{\includegraphics[trim = {0cm, 2.4cm, 0cm, 0cm}, clip=True,width=1\textwidth]{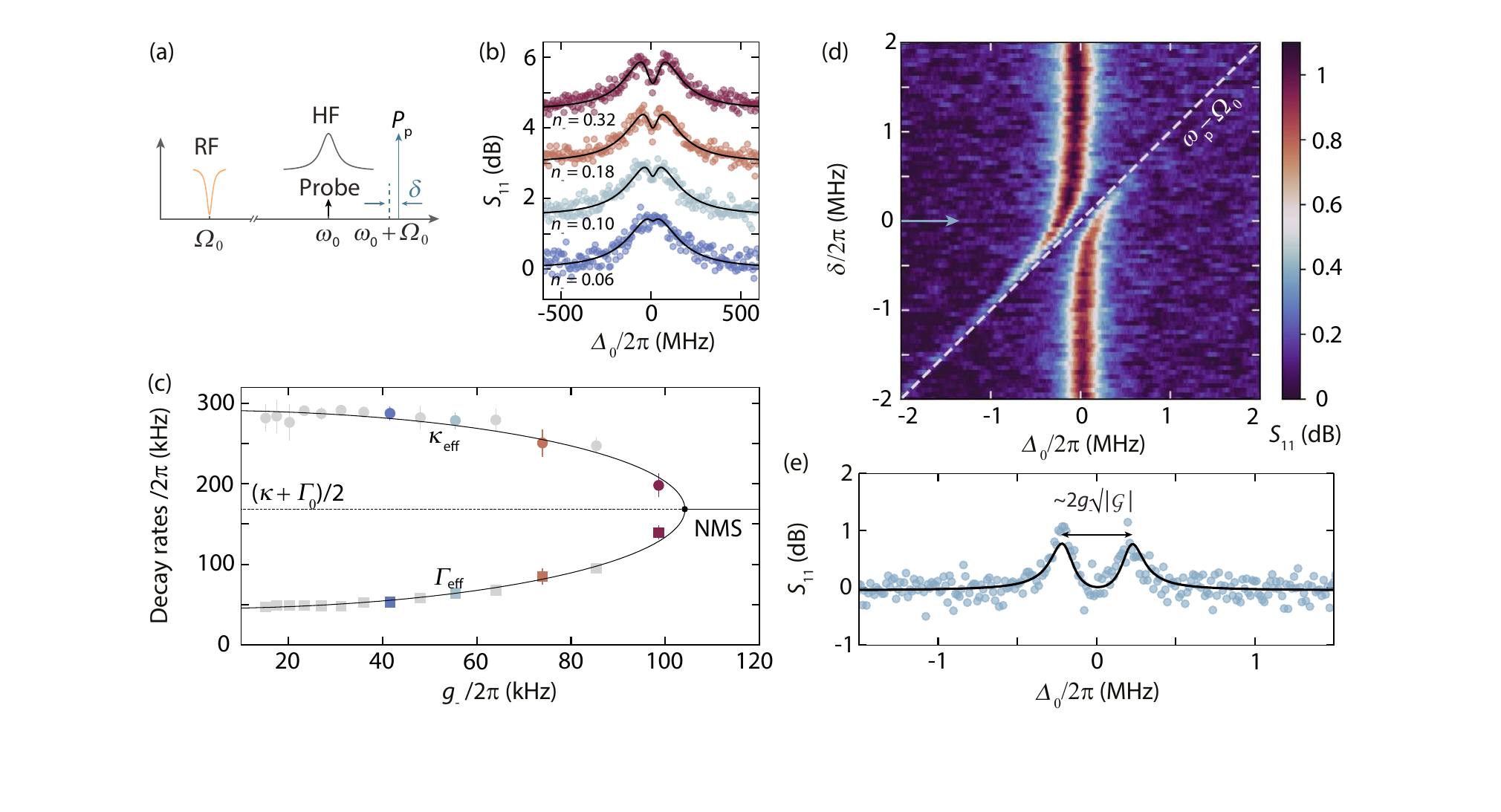}}
	\caption{\textsf{\textbf{Dynamical backaction inversion and normal-mode splitting with an effective negative mass microwave mode.} (a) Schematic of the experiment. A pump tone (power $P_\mathrm{p}$) is applied at $\omega_\mathrm{p} = \omega_0 + \varOmega_0 + \delta$. A small probe field tracks $S_{11}$ around $\omega_0$. (b) $S_{11}$ vs $\varDelta_0 = \omega - \omega_0$ for different $P_\mathrm{p}$ and $\delta = 0$. The cavity resonance peak displays photon-pressure induced absorption, an interference effect equivalent to optomechanically induced absorption. From the fits (lines), we obtain $\varGamma_\mathrm{eff}$ and $\kappa_\mathrm{eff}$, respectively, which are plotted in panel (c) vs $g_- = \sqrt{n_-}g_0$ with $n_-$ the intracavity pump photon number. Details can be found in \cite{Supplement}. Symbols are data, lines are following Eq.~(\ref{eqn:NMS}). The onset of normal-mode splitting is labelled with NMS. (d) $S_\mathrm{11}$ vs $\varDelta_0 = \omega - \omega_0$ and $\delta$. The arrow indicates the linescan shown in (e). The splitting between the modes for $\delta = 0$ is $\sim2\sqrt{\left|\mathcal{G}\right|}|g_-| \approx 2\pi\cdot 500\,$kHz$\,> \kappa, \varGamma_0$ and indicates the strong-coupling regime. For panels (d), (e) the HF cavity was driven with slightly different parameters and $\mathcal{G} = -0.21$ \cite{Supplement}.}}
	\label{fig:Fig2}
\end{figure*}

Without any particular measures, the HF cavity displays $\mathcal{G} = 1$.
To obtain $\mathcal{G} < 0$ we use the HF mode Kerr nonlinearity.
A strong near-resonant drive tone leads to the appearance of two quasi-modes in the probe response of the system \cite{Ochs21, FaniSani21}.
The response of one of the modes is equivalent to $\mathcal{G} > 1$ (the signal mode at $\omega_\mathrm{s}$) and the second mode shows $\mathcal{G} < 0$ (the idler mode at $\omega_0$), cf. Fig.~\ref{fig:Fig1}(c), (d).
The idler mode is closely related to Bogoliubov ghost branches observed in condensates and quantum fluids \cite{Ciuti01, Vogels02, Kohnle11, Pieczarka15, Claude22}, and so another suitable name for it would be ghost mode.
The origin of this double-mode response is four-wave mixing and parametric amplification \cite{FaniSani21}.
The probe reflection near $\omega_0$ is given by
\begin{equation}
	S_{11}(\omega \approx \omega_0) = 1 - \kappa_\mathrm{e}\chi_\mathcal{G} = 1 - \frac{\kappa_\mathrm{e}\mathcal{G}}{\frac{\kappa}{2} + i(\omega - \omega_0)}
\end{equation}
with $\mathcal{G} = - 0.35$, $\kappa = 2\pi\cdot 290\,$kHz and $\omega_0 = 2\pi\cdot 7.211\,$GHz (the smaller $\kappa$ compared to the undriven case is most likely related to saturation of two-level systems by the drive \cite{Capelle20, Rodrigues22}).
Although both $\mathcal{G} > 1$ and $\mathcal{G}<0$ can lead to a resonance peak in the reflection response, the interpretation behind it is fundamentally different.
For $\mathcal{G} > 1$ the peak arises from amplification of the intracavity field, for $\mathcal{G} < 0$ from a phase shift of $\pi$, which is equivalent to a negative mass HO \cite{Supplement}.
\begin{figure*}
	\centerline{\includegraphics[trim = {3cm, 1.5cm, 7cm, 2cm}, clip=True,width=0.75\textwidth]{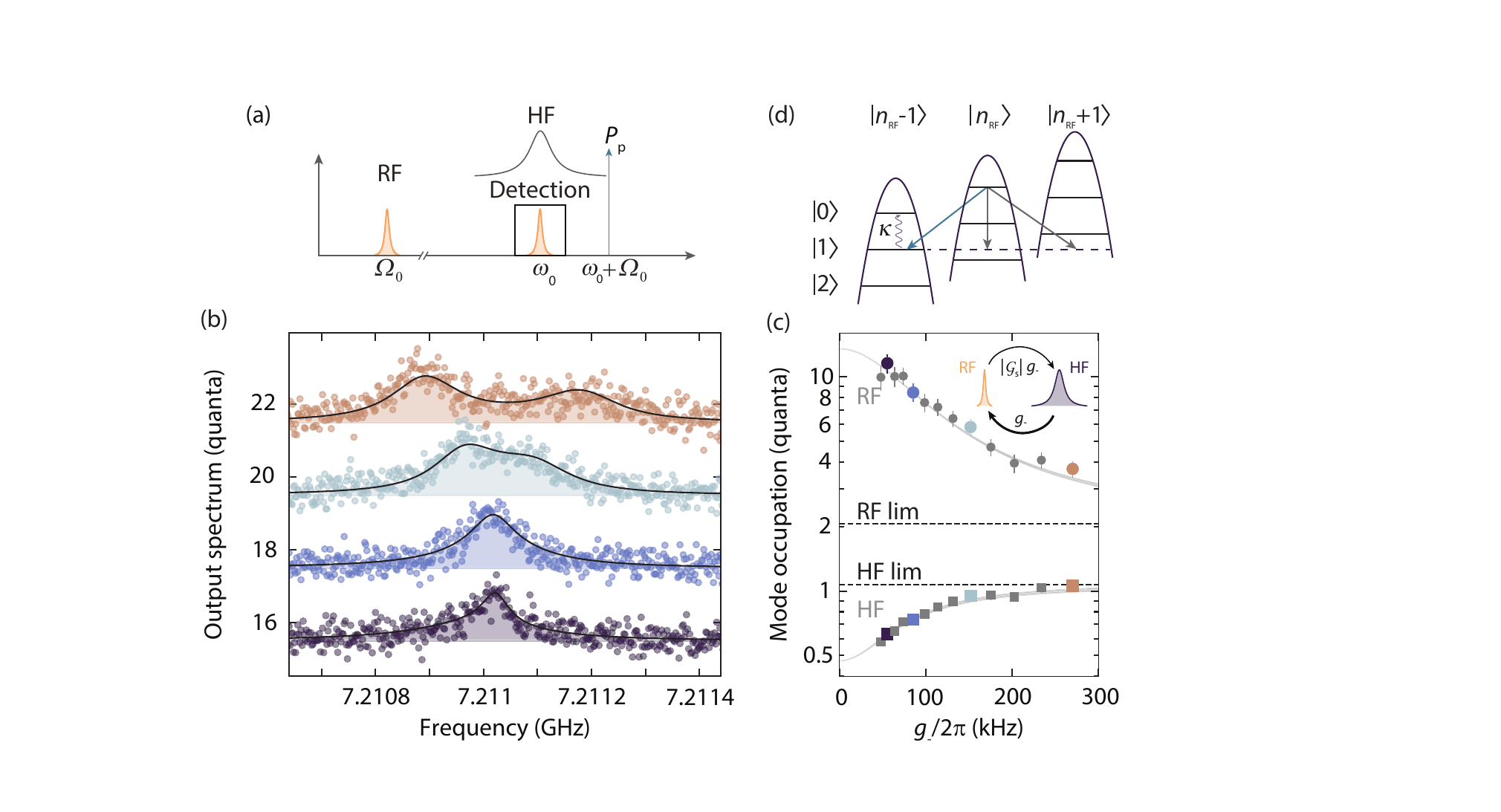}}
	\caption{\textsf{\textbf{Blue-detuned photon-pressure sideband-cooling with an effective negative mass microwave reservoir.} (a) Schematic of the experiment. A pump tone with power $P_\mathrm{p}$ is applied at $\omega_\mathrm{p} = \omega_0 + \Omega_0$. The HF mode output power spectral density (PSD) is recorded for each $P_\mathrm{p}$ with a spectrum analyzer. (b) HF mode PSD in units of quanta for various $P_\mathrm{p}$, curves are offset manually by $+2$ each (lowest curve unshifted). Circles are data, lines and shaded areas are fits. From the fits we extract the occupation of both modes $n_\mathrm{fin}^\mathrm{RF}$ and $n_\mathrm{fin}^\mathrm{HF}$, the result is plotted in panel (c) vs $g_-$, circles are data, gray lines are theory including uncertainties \cite{Supplement}. Due to $|\mathcal{G}| < 1$, we also observe an imbalance in the final mode occupations $n_\mathrm{lim}^\mathrm{RF} > n_\mathrm{lim}^\mathrm{HF}$, which is related to nonreciprocal heat flow \cite{Rodrigues22}. The two horizontal dashed lines show the limit values for the RF and HF occupations for $g_- \gg \kappa, \varGamma_0$. Error bars in the data points consider the standard deviation of $n_\mathrm{th}^\mathrm{RF}$ and the fitting error of each PSD \cite{Supplement}. (d) Energy level schematic of the effective photon scattering process that leads to blue-detuned RF mode cooling.}}
	\label{fig:Fig4}
\end{figure*}
The negative-mass-inverted susceptibility, however, is not limited to an apparent effect in the reflection response to a probe tone.
Sideband fields from an additional photon-pressure pump tone will also experience the effective inversion.
In the following we will consider the driven system with an additional photon-pressure pump tone applied at the blue idler-mode sideband $\omega_\mathrm{p} = \omega_0 + \varOmega_0$, cf. Fig.~\ref{fig:Fig2}(a).
We will also work in a reduced HF mode space, where all we consider is a single generalized mode with $\mathcal{G} < 0$.
The linearized and approximated equations of motion in Fourier space for the intracavity field fluctuation $\hat{c}, \hat{c}^\dagger$ and the RF mode amplitude $\hat{b}, \hat{b}^\dagger$ ladder operators then are \cite{Rodrigues22, Supplement}
\begin{eqnarray}
	\frac{\hat{c}}{\chi_\mathcal{G}} & = & -ig_-\hat{b}^\dagger + \sqrt{\kappa_\mathrm{e}}\hat{\xi}_\mathrm{e, eff} + \sqrt{\kappa_\mathrm{i}}\hat{\xi}_\mathrm{i, eff} \label{eqn:EOM1}\\
	\frac{\hat{b}^\dagger}{\overline{\chi}_0} & = & ig_-^*\hat{c} + \sqrt{\varGamma_\mathrm{e}}\hat{\zeta}_\mathrm{e}^\dagger + \sqrt{\varGamma_\mathrm{i}}\hat{\zeta}_\mathrm{i}^\dagger \label{eqn:EOM2}
\end{eqnarray}
where $\overline{\chi}_0^{-1} = \varGamma_0/2 + i(\varOmega + \varOmega_0)$ is the RF mode susceptibility, $\varOmega = \omega - \omega_\mathrm{p}$ is the frequency relative to the pump tone, $\kappa_\mathrm{e}, \kappa_\mathrm{i}$ and $\varGamma_\mathrm{e}, \varGamma_\mathrm{i}$ are the external and internal coupling rates of the circuits to their corresponding baths and $\hat{\zeta}_\mathrm{e, i}$ are the noise input operators for the RF mode following $\left\langle \hat{\zeta}_\mathrm{e, i}^\dagger\hat{\zeta}_\mathrm{e, i} \right\rangle = n_\mathrm{e, i}^\mathrm{RF}$.
The multi-photon coupling rate is $g_- = \gamma_-g_0$ with the sideband pump intracavity amplitude $\gamma_-$, related to the intracavity pump photon number $n_-$ through $n_- = |\gamma_-|^2$ \cite{Footnote1}.
We combine Eqs.~(\ref{eqn:EOM1}, \ref{eqn:EOM2}) to find the effective RF mode susceptibility and what we get looks formally identical to the usual blue sideband pumped system with
\begin{equation}
	\overline{\chi}_0^\mathrm{eff}(\varOmega) = \frac{1}{\frac{\varGamma_0}{2} + i\left(\varOmega + \varOmega_0\right) - |g_-|^2\chi_\mathcal{G}(\varOmega)}.
\end{equation}
When pumping exactly on the blue sideband, the complex eigen-frequencies $\tilde{\varOmega}_\pm$ of this susceptibility are given by
\begin{equation}
	\tilde{\varOmega}_\pm = -\varOmega_0 + i\frac{\kappa + \varGamma_0}{4} \pm i\sqrt{ \left[\frac{\kappa - \varGamma_0}{4}\right]^2 + \mathcal{G}|g_-|^2},
	\label{eqn:NMS}
\end{equation}
obtained through the condition $\chi_0^\mathrm{eff}(\tilde{\varOmega})^{-1} = 0$.
As long as the coupling is not too large $\left[\frac{\kappa - \varGamma_0}{4}\right]^2 > -\mathcal{G}|g_-|^2 $, the expression under the root will be $>0$ and the two modes (RF and HF) will just display modified linewidths $\varGamma_\mathrm{eff}$ and $\kappa_\mathrm{eff}$ given by $2\mathrm{Im}(\tilde{\varOmega_\pm})$.
Hence, the effective RF linewidth $\varGamma_\mathrm{eff}$ will indeed increase with increasing $|g_-|$ in the case $\mathcal{G} < 0$; we obtain blue-sideband-pumped positive dynamical backaction damping.
The result is also in excellent quantitative agreement with the experimental data, cf. Fig.~\ref{fig:Fig2}.
The phase shift of $\pi$ in the susceptibility implies that for a blue-detuned pump the intracavity field is not adjusting in a way which amplifies the RF oscillation as in a regular cavity, but in a way that opposes the RF ``motion" and is therefore reducing its amplitude.
For even stronger pumping $-\mathcal{G}|g_-|^2 > \left[\frac{\kappa - \varGamma_0}{4}\right]^2$ the square root in Eq.~(\ref{eqn:NMS}) becomes imaginary and we find two distinct resonance frequency solutions with identical damping rates $(\kappa + \varGamma_0)/2$.
We witness the onset of normal-mode splitting, and for slightly larger $|g_-|$ even the onset of the strong-coupling regime \cite{Groeblacher09, Teufel11a, Bothner21} (splitting larger than $(\kappa + \varGamma_0)/2$), cf. Fig.~\ref{fig:Fig2}(d), (e)), which is something that could not happen with $\mathcal{G}>0$ for blue-detuned pumping.
Finally, we demonstrate that the inverted dynamical backaction also leads to blue-detuned sideband cooling of the RF mode.
For this experiment, we put again a pump to the blue sideband of the negative mass HF mode and detect the HF mode output noise with a spectrum analyzer, cf. Fig.~\ref{fig:Fig4}(a).
From the observed power spectral density (PSD), cf. Fig.~\ref{fig:Fig4}(b), we can infer the occupation of both, the HF mode and the RF mode by fitting the PSD in units of quanta \cite{Supplement}.
For this, we assume the unpumped thermal occupation of the HF mode to be negligible.
The result reveals that indeed the RF mode is sideband-cooled by the blue-detuned pump tone, cf. Fig.~\ref{fig:Fig4}(c), (d).
The starting residual occupation is $n_\mathrm{th}^\mathrm{RF} \sim 13.5$ and the cooling reduces this to $n_\mathrm{fin}^\mathrm{RF} \sim 3.5$.
At the same time, the effective HF mode occupation is increased from its effective occupation without the cooling tone $\tilde{n}_\mathrm{th}^\mathrm{HF} \sim 0.5$ to $n_\mathrm{fin}^\mathrm{HF} \sim 1$.
For the limit $g_- \rightarrow \infty$ we get $n_\mathrm{lim}^\mathrm{RF} > n_\mathrm{lim}^\mathrm{HF}$ and not $n_\mathrm{lim}^\mathrm{RF} = n_\mathrm{lim}^\mathrm{HF}$ as expected for standard photon-pressure and optomechanical systems \cite{Dobrindt08, Teufel11, Rodrigues21}.
The reason behind this asymmetry is nonreciprocal heat transfer due to $|\mathcal{G}| \neq 1$ as also discussed in Ref.~\cite{Rodrigues22}.
There is a simple level-diagram interpretation of this blue-detuned cooling with an effective negative mass reservoir.
In a negative mass mode adding one excitation corresponds to lowering the energy by $\hbar\omega_0$ \cite{Glauber86}.
The sideband-transition $|n_\mathrm{HF}, n_\mathrm{RF} \rangle \rightarrow |n_\mathrm{HF}+1, n_\mathrm{RF}-1 \rangle$, which leads to cooling of the RF mode, has the energy $\hbar(\omega_0 + \Omega_0)$, cf. Fig.~\ref{fig:Fig4}(d), which corresponds to a blue-detuned sideband photon.
In that sense, the blue-detuned cooling actually serves as a probe for the sign of the effective oscillator mass.
With a positive mass the blue-detuned pump would correspond to a two-mode-squeezing interaction and amplification instead of cooling, since the down-scattered photons from the HF pump would lose $\hbar\Omega_0$, and that energy would be added to the RF mode \cite{Massel11}.
With a negative mass, creation and annihilation operators in a sense swap roles and the usual TMS interaction is converted to a beam-splitter term.
Finally, we also note that the effect we report here is different from the seemingly similar effect reported recently in an optomechanical system \cite{Bothner22}, where the blue-detuned cooling is a consequence of the interference of many mechanical sidebands, while here it is a single sideband effect.
In conclusion, we have reported the engineering of an effective negative mass HO in a microwave LC circuit and demonstrated its effect to the photon-pressure coupling between this cavity and a radio-frequency LC circuit.
The effective negative-mass dynamics emerged from a combination of a Kerr nonlinearity and strong driving, a method compatible with all kinds of nonlinear oscillators.
We found that a blue-detuned sideband pump field leads to positive dynamical backaction damping, normal-mode splitting and sideband cooling, all things usually associated with red-detuned pumping and a beam-splitter interaction \cite{Gigan06, Arcizet06, Teufel08, Teufel11}.
Our results demonstrate how to mimic an effective negative mass mode in a generic Kerr oscillator.
All data presented in this paper and the corresponding processing scripts used during the analysis are available on Zenodo \cite{zenodo}.
This research was supported by the Netherlands Organisation for Scientific Research (NWO) in the Innovational Research Incentives Scheme -- VIDI, project 680-47-526.
This project has received funding from the European Research Council (ERC) under the European Union's Horizon 2020 research and innovation programme (grant agreement No. 681476 - QOMD), from the European Union's Horizon 2020 research and innovation programme (grant agreement No. 732894 - HOT), and from the Deutsche Forschungsgemeinschaft (DFG, German Research Foundation) via grant No. 490939971 (BO 6068/1-1).
\let\oldaddcontentsline\addcontentsline
\renewcommand{\addcontentsline}[3]{}
\let\addcontentsline\oldaddcontentsline
\clearpage

\widetext

\begin{center}
\noindent\textbf{\large Supplemental Material for:\\ Photon-Pressure with an Effective Negative Mass Microwave Mode}

\normalsize
\vspace{.3cm}

\noindent{I.~C.~Rodrigues, G.~A.~Steele, and D.~Bothner}
\end{center}
\vspace{.2cm}

\renewcommand{\theequation}{S\arabic{equation}}

\renewcommand{\bibnumfmt}[1]{[S#1]}
\renewcommand{\citenumfont}[1]{S#1}

\setcounter{figure}{0}
\setcounter{equation}{0}

\addtocontents{toc}{\protect\setcounter{tocdepth}{0}}

\tableofcontents

\newpage

\renewcommand{\figurename}{Supplemental Figure}

\section{Supplemental Note 1: Theory}
\subsection{Classical description of a negative-mass mechanical oscillator}
If we have a point-like object with a negative inertial mass $m_- = -m_+$, where $m_+$ is the positive mass of an equivalent standard object, then Newton's second law of motion becomes
\begin{equation}
	F = -m_+\ddot{x}
\end{equation}
with the force $F$ and the position $x$.
What do we have to do now if we want to construct an harmonic oscillator with such a mass?
First, we need to put it into a parabolic potential as usual, but one that is inverted, i.e., one in which the energy gets smaller with increasing $x$.
If we were to choose a standard potential, we would get
\begin{equation}
	-m_+ \ddot{x} + k_+ x = F_\mathrm{ex}
\end{equation}
with external forces being summarized in $F_\mathrm{ex}$ and the spring constant $k_+$ describing the restoring force in the parabolic potential.
Such an equation does not lead to a periodically oscillating motion without external force and to no amplitude enhancement at a specific frequency, which is also known as resonance.
To actually obtain a harmonic oscillator, we need a negative spring constant $k_- = -k_+$, which leads to the equation of motion
\begin{equation}
	-m_+ \ddot{x} - k_+ x = F_\mathrm{ex}.
\end{equation}
How can we understand this intuitively?
The characteristic of a harmonic oscillator is that when the particle moves out of the equilibrum position, a linear-in-displacement force is pulling it back to the equilibrium position. 
A positive-mass particle responds with moving towards the direction of this restoring force.
A negative-mass particle though moves opposite to the restoring force, i.e., if we use a standard spring, the back-pulling force will push the particle further away from its equilibrium position.
Since the force of a spring is getting stronger with increasing displacement, this immediately leads to an instability.
With similar considerations, we find that the friction force in a stable negative-mass oscillator needs to act opposite to the usual friction force.
This means it will add energy instead of removing it.
Otherwise the oscillator will again not be stable and its amplitude will exponentially grow with time.
The resulting equation of motion is
\begin{equation}
	-m_+ \ddot{x} - \gamma_+\dot{x} - k_+ x = F_\mathrm{ex}.
\end{equation}
that we can also write as
\begin{equation}
	\ddot{x} + \kappa\dot{x} +\omega_0^2 x = -\frac{F_\mathrm{ex}}{m_+}
	\label{eqn:EOMx}
\end{equation}
with $\kappa = \gamma_+/m_+$ and $\omega_0 = \sqrt{k_+/m_+}$ both being exactly the same as in the positive-mass case and in particular $>0$.
The only difference between a positive and a negative mass oscillator in their equations of motion is therefore a minus sign on the right-hand-side.
For the possible solutions of this equation of motion, this minus sign is therefore equivalent to a phase shift of $\pi$.
Note that this is the only dynamical difference, if we were to measure for instance the displacement.
By Fourier transform we can also solve the equation of motion and get
\begin{equation}
	-\omega^2 x(\omega) + i\omega\kappa x(\omega) + \omega_0^2 x(\omega) = -\frac{F_\mathrm{ex}(\omega)}{m_+}
\end{equation}
or
\begin{equation}
	x(\omega) = \chi_-(\omega)F_\mathrm{ex}(\omega), ~~~~~ \chi_-(\omega) = -\frac{1}{m_+}\frac{1}{\omega_0^2 - \omega^2 + i\omega\kappa} = -\chi_+(\omega)
\end{equation}
or in high-$Q$ approximation
\begin{equation}
	\chi_-(\omega) \approx -\frac{1}{2im_+\omega_0} \frac{1}{\frac{\kappa}{2} + i(\omega - \omega_0)}.
\end{equation}
In both cases, the difference in susceptibility and response, respectively, is a phase shift of $\pi$.
We can also find the classical equation of motion for the complex amplitude $\alpha$, which is defined via
\begin{equation}
	\alpha = \dot{x} + i\omega_1 x
\end{equation}
and which (except for the normalization) is the classical equivalent to the quantum annihilation operator $\hat{a}$.
Here,
\begin{eqnarray}
	\omega_1 & = & \omega_0\left[\left( 1 - \frac{1}{4Q^2} \right)^{1/2} - i\frac{1}{2Q} \right] \\
	& = & \tilde{\omega}_0 - i\frac{\kappa}{2}
\end{eqnarray}
is the complex-valued resonance frequency, whose real part $\tilde{\omega}_0$ is slightly shifted compared to $\omega_0$ by the damping and whose imaginary part is half the decay rate.
After insertion into Eq.~(\ref{eqn:EOMx}) and some algebraic manipulations we get two new equations of motion for $\alpha$ and $\alpha^*$
\begin{eqnarray}
	\dot{\alpha} + \left(\frac{\kappa}{2}  - i\tilde{\omega}_0\right)\alpha & = &  \pm \frac{F_\mathrm{ex}}{m_+} \\
	\dot{\alpha}^* + \left(\frac{\kappa}{2}  + i\tilde{\omega}_0\right)\alpha^* & = &  \pm \frac{F_\mathrm{ex}}{m_+}
\end{eqnarray}
where the $\pm$ on the right hand side represents the sign for positive and negative mass, respectively.
We now discuss the total energy of the negative mass oscillator.
The potential energy can be written straightforwardly from the negative potential as
\begin{equation}
	E_\mathrm{pot} = -\frac{1}{2}m_+\omega_0^2 x^2.
\end{equation}
Furthermore, the kinetic energy is given by
\begin{equation}
	E_\mathrm{kin} = -\frac{1}{2}m_+\dot{x}^2.
\end{equation}
With those two at hand, we can write down the classical Lagrangian
\begin{equation}
	\mathcal{L} = -\frac{1}{2}m_+\dot{x}^2 +\frac{1}{2}m_+\omega_0^2 x^2
\end{equation}
and find
\begin{equation}
	p = \frac{\partial \mathcal{L}}{\partial \dot{x}} = -m_+\dot{x} = -p_+.
\end{equation}
We note that here the momentum changed sign compared to the positive mass case.
Finally, we write down the classical Hamiltonian as
\begin{equation}
	\mathcal{H} = -\frac{p^2}{2m_+} - \frac{1}{2}m_+\omega_0^2 x^2.
\end{equation}
which is identical to the Hamiltonian of a positive mass, but with a global change of sign from plus to minus.
We introduce now the classical amplitudes
\begin{eqnarray}
	a & = & \sqrt{\frac{m_+\omega_0}{2\hbar}}x + i\frac{p}{\sqrt{2\hbar m_+ \omega_0}} \\
	a^* & = & \sqrt{\frac{m_+\omega_0}{2\hbar}}x - i\frac{p}{\sqrt{2\hbar m_+ \omega_0}}
\end{eqnarray}
with real-valued $x,p$.
This relation can be inverted to give
\begin{eqnarray}
	x & = & \sqrt{\frac{\hbar}{2m_+\omega_0}}(a + a^*) \\
	p & = & \frac{1}{i}\sqrt{\frac{\hbar m_+ \omega_0}{2}}(a-a^*)
\end{eqnarray}
and after injection into the Hamiltonian we get
\begin{equation}
	\mathcal{H} = -\hbar\omega_0a^*a.
\end{equation}
Note that $a$ is essentially the same as $\alpha$, just with a different normalization and without taking any damping into account.
It is furthermore interesting to note that since $p = -p_+$ the fields $a$ and $a^*$ swap their places compared to the positive mass case.
This can most likely be interpreted as a signature that a blue-detuned two-mode squeezing term $\hat{a}\hat{b} + \hat{a}^\dagger \hat{b}^\dagger$ in case of a positive oscillator turns into a beam-splitter term $\hat{a}^\dagger\hat{b} + \hat{a} \hat{b}^\dagger$ if the mass is inverted.
The considerations we presented and discussed in this section demonstrate that for the classical case, a negative mass oscillator is fully characterized by a phase shift of $\pi$ in its response.
Since Hamiltonian equations and equations of motion have the same amount of information on the dynamical properties of a system, a phase shift of $\pi$ is therefore equivalent to a minus sign in the Hamiltonian. In other words, it would be equivalent to a negative energy ladder and to a negative effective mass, respectively.

\subsection{A negative capacitance RLC circuit and its reflection coefficient}
Since we are dealing with LC circuits in this work, we also briefly discuss the circuit treatment of a negative capacitance version, which is coupled to a usual transmission line.
Since the equations of motion would be completely equivalent to the mechanical oscillator above, we move along a slightly different formal path though.
Since we know already that for a stable oscillator, we need, $R_-, L_-, C_- < 0$, we start with considering the input impedance of a negative capacitance, parallel RLC circuit
\begin{eqnarray}
	Z_\mathrm{in}^\mathrm{RLC} & = & \left( -\frac{1}{R} - \frac{1}{i\omega L} - i\omega C\right)^{-1} \\
	& = & - \frac{R}{1 + \frac{R}{i\omega L} + i\omega RC}
\end{eqnarray}
where $R, L, C >0$.
Next, we add a coupling capacitance $C_\mathrm{c}$ in series and get
\begin{equation}
	Z_\mathrm{in} = -\frac{R}{1 + \frac{R}{i\omega L} + i\omega RC} - \frac{1}{i\omega C_\mathrm{c}}.
\end{equation}
For frequencies close to the resonance frequency $\omega_0 = 1/\sqrt{L\left( C + C_\mathrm{c} \right)}$, the input impedance can be approximated as
\begin{equation}
	Z_\mathrm{in} \approx -\frac{Z_0}{\kappa_\mathrm{e}}\left(\kappa_\mathrm{i} +2i\varDelta \right)
\end{equation}
where $\varDelta = \omega - \omega_0$ and
\begin{equation}
	\kappa_\mathrm{i} = \frac{1}{R\left(C + C_c\right)}, ~~~~~ \kappa_\mathrm{e} = \frac{Z_0 C_c^2}{L\left(C + C_c\right)^2}
\end{equation}
the internal and external decay rates, respectively, and $\kappa_\mathrm{i}, \kappa_\mathrm{e} > 0$.
The reflection coefficient of the circuit when connected to a feedline with characteristic impedance $Z_0$ is then 
\begin{eqnarray}
	S_{11} & = & \frac{Z_\mathrm{in} - Z_0}{Z_\mathrm{in} + Z_0} \nonumber \\
	& = & \frac{\kappa_\mathrm{i} + 2i\varDelta + \kappa_\mathrm{e}}{\kappa_\mathrm{i} + 2i\varDelta - \kappa_\mathrm{e}} \nonumber \\
	& = & 1 + \frac{2\kappa_\mathrm{e}}{\kappa + 2i\varDelta}
\end{eqnarray}
with $\kappa = \kappa_\mathrm{i} - \kappa_\mathrm{e}$.
Hence, as long as $\kappa_\mathrm{e} < \kappa_\mathrm{e}$, we get exactly the phenomenology we observe in this paper for the ghost mode, i.e., a peak in reflection and an inverted phase response.
Note that the resonance-singularity for $\kappa_\mathrm{e} = \kappa_\mathrm{i}$ could only occur, if we had built a real negative mass oscillator with a negative friction force, since then the "losses" to the feedline would act as a permanent source of circuit driving and could cmpensate the losses (=energy increase) in the circuit itself.
This also shows one of the limitations of our effective negative mass picture, since of course we do not have a truly negative resistor in the mode.

\subsection{Device equations of motion}
The linearized and Fourier transformed Heisenberg-Langevin equations of motion for the RF and HF intracavity fields including a parametric drive, a photon-pressure sideband-pump and fluctuation (= noise and probe) input fields are given by
\begin{eqnarray}
	\frac{\hat{b}_0}{\chi_{0, 0}} & = & -i\left(g_-^*\hat{c}_0 + g_-\hat{c}_0^\dagger \right) - ig_\alpha\left(\hat{c}_1 + \hat{c}_1^\dagger \right) - i\left(g_+^*\hat{c}_2 + g_+\hat{c}_2^\dagger \right) + \sqrt{\varGamma_\mathrm{e}}\hat{\zeta}_\mathrm{e, 0} + \sqrt{\varGamma_\mathrm{i}}\hat{\zeta}_\mathrm{i, 0} \\
	\frac{\hat{c}_0}{\chi_{\mathrm{p}, 0}} & = & -ig_-\left(\hat{b}_0 + \hat{b}_0^\dagger \right) - ig_\alpha\left(\hat{b}_{-1} + \hat{b}_1^\dagger \right) - ig_+\left(\hat{b}_{-2} + \hat{b}_2^\dagger \right) + i\mathcal{K}n_\mathrm{d}\hat{c}_2^\dagger + \sqrt{\kappa_\mathrm{e}}\hat{\xi}_\mathrm{e, 0} + \sqrt{\kappa_\mathrm{i}}\hat{\xi}_\mathrm{i, 0} + i\sqrt{\kappa_\mathrm{e}}\hat{c}_\mathrm{in, 0},
\end{eqnarray}
where
\begin{equation}
	\chi_{0, 0} = \frac{1}{\frac{\varGamma_0}{2} + i(\varOmega - \varOmega_0)}, ~~~ \chi_\mathrm{p, 0} = \frac{1}{\frac{\kappa}{2} + i\left(\varDelta_\mathrm{p} - 2\mathcal{K}n_\mathrm{d} + \varOmega\right)}.
\end{equation}
For a very detailed derivation and the step-by-step linearization see the Material of Ref.~\cite{Rodrigues22_SI}.
In these equations, the annihilation (creation) operators for the radio-frequency mode are denoted with $\hat{b}$ ($\hat{b}^\dagger$) and for the high-frequency cavity with $\hat{c}$ ($\hat{c}^\dagger$).
The indices at the individual Fourier components of the fields and susceptibilites $j$ are defined as $\hat{b}_j = \hat{b}(\varOmega + j\varOmega_\mathrm{dp})$ and $\hat{b}^\dagger_j = \hat{b}^\dagger(-\varOmega + j\varOmega_\mathrm{dp})$ with the drive-pump detuning $\varOmega_\mathrm{dp} = \omega_\mathrm{d} - \omega_\mathrm{p}$.
The resonance frequency and damping rate of the RF mode are given by $\varOmega_0$ and $\varGamma_0$, respectively, and the damping rate is the sum of the internal and external damping rates $\varGamma_\mathrm{i}$ and $\varGamma_\mathrm{e}$ with $\varGamma_0 = \varGamma_\mathrm{i} + \varGamma_\mathrm{e}$.
For the undriven HF mode, resonance frequency and damping rates are given by $\omega_\mathrm{c}$ and $\kappa = \kappa_\mathrm{i} + \kappa_\mathrm{e}$.
The photon-pressure coupling rates $g_-, g_+, g_\alpha$ are given by $\gamma_- g_0, \gamma_+g_0$, and $\alpha_\mathrm{d}g_0$, where $\alpha_\mathrm{d}$ is the intracavity field at the parametric drive frequency, $\gamma_-$ is the complex intracavity field amplitude at the frequency of the photon-pressure sideband-pump and $\gamma_+$ is the corresponding field amplitude of the four-wave-mixed mirror tone of $\gamma_-$.
The photon-pressure single-photon coupling rate is $g_0$ (including possible Kerr modulation contribution, cf. Ref.~\cite{Rodrigues22_SI}), the HF cavity Kerr nonlinearity due to the Josephson inductance is $\mathcal{K}$, the detuning between photon-pressure sideband pump frequency and undriven HF resonance frequency is $\varDelta_\mathrm{p} = \omega_\mathrm{p} - \omega_\mathrm{c}$, and the intracavity drive photon number is $n_\mathrm{d} = \alpha_\mathrm{d}^2$.
For the input noise, we use $\hat{\zeta}_\mathrm{i}$ and $\hat{\zeta}_\mathrm{e}$ for the internal and external RF oscillator baths and $\hat{\xi}_\mathrm{i}, \hat{\xi}_\mathrm{e}$ for the internal and external HF cavity baths, respectively.
The correlators for the input noise operators (in the frequency domain) are given by
\begin{eqnarray}
	\langle\hat{\zeta}^\dagger(\varOmega)\hat{\zeta}(\varOmega')\rangle & = & n_\mathrm{th}^\mathrm{RF}\delta\left(\varOmega - \varOmega'\right) \nonumber \\ \langle\hat{\zeta}(\varOmega)\hat{\zeta}^\dagger(\varOmega')\rangle & = & \left(n_\mathrm{th}^\mathrm{RF} + 1\right)\delta\left(\varOmega - \varOmega'\right) \nonumber \\
	\langle\hat{\xi}^\dagger(\omega)\hat{\xi}(\omega')\rangle & = & n_\mathrm{th}^\mathrm{HF}\delta\left(\omega - \omega'\right) \nonumber \\ \langle\hat{\xi}(\omega)\hat{\xi}^\dagger(\omega')\rangle &  = & \left(n_\mathrm{th}^\mathrm{HF} + 1\right)\delta\left(\omega - \omega'\right)
\end{eqnarray}
where $n_\mathrm{th}^\mathrm{RF}$ and $n_\mathrm{th}^\mathrm{HF}$ are the thermal bath photon occupations of the RF mode and the HF cavity bath, respectively.
Note that for simplicity we omitted the subscripts for internal and external baths here, those will be used later again where necessary.
A coherent high-frequency input field such as a probe tone is finally denoted with $\hat{c}_\mathrm{in}$.
\subsection{Simplified equations of motions without noise}
First, we neglect the input noise terms and therefore get
\begin{eqnarray}
	\frac{\hat{b}_0}{\chi_{0, 0}} & = & -i\left(g_-^*\hat{c}_0 + g_-\hat{c}_0^\dagger \right) - ig_\alpha\left(\hat{c}_1 + \hat{c}_1^\dagger \right) - i\left(g_+^*\hat{c}_2 + g_+\hat{c}_2^\dagger \right) \\
	\frac{\hat{c}_0}{\chi_{\mathrm{p}, 0}} & = & -ig_-\left(\hat{b}_0 + \hat{b}_0^\dagger \right) - ig_\alpha\left(\hat{b}_{-1} + \hat{b}_1^\dagger \right) - ig_+\left(\hat{b}_{-2} + \hat{b}_2^\dagger \right) + i\mathcal{K}n_\mathrm{d}\hat{c}_2^\dagger + i\sqrt{\kappa_\mathrm{e}}\hat{c}_\mathrm{in, 0}.
\end{eqnarray}
Next, we assume blue-sideband idler-mode pumping and omit all non-resonant RF terms, assuming $\varOmega \sim -\varOmega_0$, and get for the cavity
\begin{eqnarray}
	\frac{\hat{c}_0}{\chi_{\mathrm{p}, 0}} & = & -ig_- \hat{b}_0^\dagger + i\mathcal{K}n_\mathrm{d}\hat{c}_2^\dagger + i\sqrt{\kappa_\mathrm{e}}\hat{c}_\mathrm{in, 0} \nonumber \\
	\frac{\hat{c}_2^\dagger}{\overline{\chi}_{\mathrm{p}, 2}} & = & ig_+^* \hat{b}_0^\dagger - i\mathcal{K}n_\mathrm{d}\hat{c}_0 - i\sqrt{\kappa_\mathrm{e}}\hat{c}_\mathrm{in, 2}^\dagger.
\end{eqnarray}
With $\hat{c}_\mathrm{in, 2}^\dagger = 0$ and by eliminating $\hat{c}_2^\dagger$ from the first equation using the second, we obtain
\begin{eqnarray}
	\frac{\hat{c}}{\chi_{\mathrm{g}}} & = & -i\left[g_- - i\mathcal{K}n_\mathrm{d}\overline{\chi}_\mathrm{p}g_+^*\right] \hat{b}^\dagger + i\sqrt{\kappa_\mathrm{e}}\hat{c}_\mathrm{in},
\end{eqnarray}
where we also dropped the subscripts "0" and "2".
The new susceptibility here is given by
\begin{equation}
	\chi_\mathrm{g} = \frac{\chi_\mathrm{p, 0}}{1 - \mathcal{K}^2 n_\mathrm{d}^2 \chi_\mathrm{p, 0}\overline{\chi}_\mathrm{p, 2}} = \frac{\chi_\mathrm{p}}{1 - \mathcal{K}^2 n_\mathrm{d}^2\chi_\mathrm{p}\overline{\chi}_\mathrm{p}}
	\label{eqn:FullKerr}
\end{equation}
From here, we use for our regime $g_- - i\mathcal{K}n_\mathrm{d}\overline{\chi}_\mathrm{p} \approx g_-$ and get
\begin{eqnarray}
	\frac{\hat{c}}{\chi_{\mathrm{g}}} & = & -ig_- \hat{b}^\dagger + i\sqrt{\kappa_\mathrm{e}}\hat{c}_\mathrm{in}.
\end{eqnarray}
For the RF mode we get on the other hand
\begin{equation}
	\frac{\hat{b}^\dagger}{\overline{\chi}_0} = ig_-^*\hat{c}
\end{equation}
with
\begin{equation}
	\overline{\chi}_0 = \frac{1}{\frac{\varGamma_0}{2} + i\left(\varOmega + \varOmega_0\right)}
\end{equation}
Note that these equations are equivalent to the equations of a usual radiation-pressure system with blue sideband pumping.
The only difference is the modified HF cavity susceptibility $\chi_\mathrm{g}$.

\subsection{Two-tone response of the HF cavity}

We use now that close to the idler mode (= negative mass mode), the total susceptibility can be well approximated by $\chi_\mathrm{g} \approx \mathcal{G}_\mathrm{i}\chi_\mathrm{i} = \mathcal{G}\chi_\mathrm{c}$ where
\begin{equation}
	\mathcal{G} \approx \frac{\varOmega_\mathrm{i} - \varDelta_\mathrm{d} + 2\mathcal{K}n_\mathrm{d}}{2\varOmega_\mathrm{i}}, ~~~ \varDelta_\mathrm{d} = \omega_\mathrm{d} - \omega_\mathrm{c}, ~~~ \varOmega_\mathrm{i} = \omega_0 - \omega_\mathrm{d}.
	\label{eqn:G}
\end{equation}
Here, $\mathcal{G} < 0$ is a real-valued number and $\chi_\mathrm{c}^{-1} = \kappa/2 + i(\varDelta + \varOmega)$ is a usual susceptibility ($\varDelta = \omega_\mathrm{p} - \omega_0$ is the detuning of the sideband pump from the idler resonance frequency and $\varOmega = \omega - \omega_\mathrm{p}$), which then leads to
\begin{eqnarray}
	\hat{c} & = & -ig_-\mathcal{G}\chi_\mathrm{c}\hat{b}^\dagger + i\mathcal{G}\chi_\mathrm{c}\sqrt{\kappa_\mathrm{e}}\hat{c}_\mathrm{in} \\
	& = & |g_-|^2\mathcal{G}\chi_\mathrm{c}\overline{\chi}_0 \hat{c} + i\mathcal{G}\chi_\mathrm{c}\sqrt{\kappa_\mathrm{e}}\hat{c}_\mathrm{in}
\end{eqnarray}
or
\begin{equation}
	\hat{c} = i\frac{\mathcal{G}\chi_\mathrm{c}}{1 - |g_-|^2\mathcal{G}\chi_\mathrm{c}\overline{\chi}_0}\sqrt{\kappa_\mathrm{e}}\hat{c}_\mathrm{in}.
\end{equation}
For the output field, we get
\begin{eqnarray}
	\hat{c}_\mathrm{out} & = & \hat{c}_\mathrm{in} + i\sqrt{\kappa_\mathrm{e}}\hat{c} \nonumber \\
	& = & \left( 1 - \kappa_\mathrm{e}\frac{\mathcal{G}\chi_\mathrm{c}}{1 - |g_-|^2\mathcal{G}\chi_\mathrm{c}\overline{\chi}_0} \right)\hat{c}_\mathrm{in}
\end{eqnarray}
and hence for the reflection
\begin{eqnarray}
	S_{11} & = & 1 - \kappa_\mathrm{e}\frac{\mathcal{G}\chi_\mathrm{c}}{1 - |g_-|^2\mathcal{G}\chi_\mathrm{c}\overline{\chi}_0}.
\end{eqnarray}
If we define the effective susceptibility as $\chi_\mathcal{G} = \mathcal{G}\chi_\mathrm{c}$, we get fully equivalent equations as for a simple standard optomechanical or photon-pressure system, both, for vanishing coupling $g_- = 0$
\begin{eqnarray}
	\hat{c} & = & i\chi_\mathcal{G}\sqrt{\kappa_\mathrm{e}}\hat{c}_\mathrm{in} \\
	S_{11} & = & 1 - \kappa_\mathrm{e}\chi_\mathcal{G} 
\end{eqnarray}
and for the coupled system $|g_-| > 0$
\begin{eqnarray}
	\hat{c} & = & i\frac{\chi_\mathcal{G}}{1 - |g_-|^2\chi_\mathcal{G}\overline{\chi}_0}\sqrt{\kappa_\mathrm{e}}\hat{c}_\mathrm{in} \\
	S_{11} & = & 1 - \kappa_\mathrm{e}\frac{\chi_\mathcal{G}}{1 - |g_-|^2\chi_\mathcal{G}\overline{\chi}_0}.
\end{eqnarray}
The latter relations we can also write as
\begin{eqnarray}
	\hat{c} & = & i\chi_\mathcal{G}^\mathrm{eff}\sqrt{\kappa_\mathrm{e}}\hat{c}_\mathrm{in} \nonumber \\
	S_{11} & = & 1 - \kappa_\mathrm{e}\chi_\mathcal{G}^\mathrm{eff}
\end{eqnarray}
when we define
\begin{equation}
	\chi_\mathcal{G}^\mathrm{eff} = \frac{\mathcal{G}\chi_\mathrm{c}}{1 - |g_-|^2\mathcal{G}\chi_\mathrm{c}\overline{\chi}_0}.
	\label{eqn:chiGeff_orig}
\end{equation}
\subsection{Expressions for the intracavity gain $\mathcal{G}$}
As detailed in the Supplementary Material of Ref.~\cite{Rodrigues22_SI}, the gain close to the idler mode frequency $\omega_0$ can be approximated as
\begin{equation}
	\mathcal{G} = \frac{1}{\overline{\chi}_\mathrm{p}(\varOmega_\mathrm{i})\left(\kappa + 2i\varOmega_\mathrm{i}\right)} \approx \frac{\varOmega_\mathrm{i} - \varDelta_\mathrm{d} + 2\mathcal{K}n_\mathrm{d}}{2\varOmega_\mathrm{i}}
\end{equation}
where $\varOmega_\mathrm{i} = \omega_0 - \omega_\mathrm{d}$ is the idler mode resonance frequency with respect to the parametric drive and given by
\begin{equation}
	\varOmega_\mathrm{i} = \sqrt{\left(\varDelta_\mathrm{d} - \mathcal{K}n_\mathrm{d}\right)\left(\varDelta_\mathrm{d} - 3\mathcal{K}n_\mathrm{d}\right)}.
\end{equation}
As it will be useful below, we give another relation
\begin{equation}
	\frac{\mathcal{G} - 1}{\mathcal{G}} \approx \mathcal{K}^2 n_\mathrm{d}^2 |\overline{\chi}_\mathrm{p}|^2
\end{equation}
valid for $\varOmega_\mathrm{i} \gg \kappa$.

\subsection{Dynamical backaction}
Next, we calculate the dynamical backaction to the RF mode.
For the $\hat{b}^\dagger$ operator we get
\begin{equation}
	\frac{\hat{b}^\dagger}{\overline{\chi}_0} = ig_-^*\hat{c} - i\sqrt{\varGamma_\mathrm{e}}\hat{b}_\mathrm{in}^\dagger
\end{equation}
where we added an input drive on the RF circuit feedline for clarity.
With
\begin{equation}
	\hat{c}= -ig_-\chi_\mathcal{G}\hat{b}^\dagger
\end{equation}
we can also write
\begin{equation}
	\frac{\hat{b}^\dagger}{\overline{\chi}_0^\mathrm{eff}} = -i\sqrt{\varGamma_\mathrm{e}}\hat{b}_\mathrm{in}^\dagger
\end{equation}
with the effective RF susceptibility
\begin{equation}
	\overline{\chi}_0^\mathrm{eff} = \frac{1}{\frac{\varGamma_0}{2} + i\left( \varOmega + \varOmega_0 \right) - |g_-|^2\chi_\mathcal{G}}
\end{equation}
From here it is obvious that (in the weak coupling regime) the dynamical photon-pressure backaction is given by
\begin{eqnarray}
	\varGamma_\mathrm{pp} & = & -2|g_-|^2\mathrm{Re}\left( \chi_\mathcal{G} \right) \\
	\delta\varOmega_0 & = & -|g_-|^2\mathrm{Im}\left( \chi_\mathcal{G} \right)
\end{eqnarray}
or with
\begin{equation}
	\chi_\mathcal{G} = \frac{\mathcal{G}}{\frac{\kappa}{2} + i\left(\varDelta + \varOmega \right)}
\end{equation}
in the form
\begin{eqnarray}
	\varGamma_\mathrm{pp} & = & -\mathcal{G}|g_-|^2\frac{\kappa}{\frac{\kappa^2}{4} + \left(\varDelta + \varOmega\right)^2} \\
	\delta\varOmega_0 & = & \mathcal{G}|g_-|^2\frac{\varDelta + \varOmega }{\frac{\kappa^2}{4} + \left(\varDelta + \varOmega\right)^2}
\end{eqnarray}
which for $\mathcal{G} = +1$ would be just the usual dynamical backaction for blue-sideband pumping of a generalized radiation-pressure system.
For $\mathcal{G} = -1$, however, which corresponds to a negative mass or an inverted susceptibility
\begin{equation}
	\chi_\mathcal{G} = -\chi_\mathrm{c}
\end{equation}
we get an inversion of the photon-pressure backaction and blue-detuned pumping leads to optical damping and cooling.
Of course, any $\mathcal{G} < 0$ leads to the same inversion effect, but its strength is scaled with a factor $|\mathcal{G}|$.
Beyond the weak coupling regime, we have to find the complex solutions $\tilde{\varOmega}_\pm$, for which $\overline{\chi}_0^\mathrm{eff}(\tilde{\varOmega})^{-1} = 0$.
Using the detuning of a blue sideband pump from the exact sideband frequency $\delta = \omega_\mathrm{p} - \left(\omega_0 + \varOmega_0\right)$, i.e. $\varDelta = \varOmega_0 + \delta$ we find the general resonance solutions as
\begin{equation}
	\tilde{\varOmega}_\pm = -\varOmega_0 - \frac{\delta}{2} + i\frac{\kappa + \varGamma_0}{4} \pm \sqrt{-\mathcal{G}|g_-|^2 - \left[ \frac{\kappa - \varGamma_0 + 2i\delta}{4}  \right]^2}.
\end{equation}
The real parts of these solutions correspond to the resonance frequencies and the imaginary parts to half the decay rates now.
When the pump is exactly on the blue sideband, we get
\begin{equation}
	\tilde{\varOmega}_\pm = -\varOmega_0 + i\frac{\kappa + \varGamma_0}{4} \pm \sqrt{-\mathcal{G}|g_-|^2 - \left[ \frac{\kappa - \varGamma_0}{4}  \right]^2}.
\end{equation}
When the terms in the square root are $< 0$, the root is imaginary and we get two solutions with identical resonance frequency $-\varOmega_0$ but different linewidths
\begin{eqnarray}
	\varGamma_\mathrm{eff} = \frac{\kappa + \varGamma_0}{2} - \sqrt{\left[ \frac{\kappa - \varGamma_0}{2}  \right]^2 + 4\mathcal{G}|g_-|^2} \\
	\kappa_\mathrm{eff} = \frac{\kappa + \varGamma_0}{2} + \sqrt{\left[ \frac{\kappa - \varGamma_0}{2}  \right]^2 + 4\mathcal{G}|g_-|^2}.
\end{eqnarray}
For $\mathcal{G} < 0$ and
\begin{equation}
	\sqrt{\left|\mathcal{G}\right|}|g_-| >  \frac{\kappa - \varGamma_0}{4}  
\end{equation}
the expression under the square root becomes positive and we get two modes with identical linewidths $\frac{\kappa + \varGamma_0}{2}$ and resonance frequencies
\begin{equation}
	\tilde{\varOmega}_\pm = -\varOmega_0 \pm \sqrt{-\mathcal{G}|g_-|^2 - \left[ \frac{\kappa - \varGamma_0}{4}  \right]^2}.
\end{equation}
which is referred to as normal-mode splitting
Note, however, that this does not yet mean the strong coupling regime, for which we require that the splitting between the modes is larger than their linewidths, i.e., that
\begin{equation}
	\sqrt{-\mathcal{G}|g_-|^2 - \left[ \frac{\kappa - \varGamma_0}{4}  \right]^2} > \frac{\kappa + \varGamma_0}{4}.
\end{equation}
The extra factor of 2 originates from the fact that the splitting is 2 times the square root.
With knowledge of these solutions, we can also re-write the expression for $\chi_\mathcal{G}^\mathrm{eff}$ Eq.~(\ref{eqn:chiGeff_orig}) as
%d
\begin{equation}
	\chi_\mathcal{G}^\mathrm{eff} = \mathcal{G}\frac{\frac{\varGamma_0}{2} + i(\varOmega + \varOmega_0)}{\left[\frac{\varGamma_\mathrm{eff}}{2} + i(\varOmega + \varOmega_0)  \right]\left[\frac{\kappa_\mathrm{eff}}{2} + i(\varOmega + \varOmega_0)  \right]}
	\label{eqn:Fig2bfunc}
\end{equation}
if we assume exact blue-sideband pumping and being below the point of normal-mode splitting.
Hence we can use this relation to fit the reflection $S_{11}$ as in Fig.~2 of the main paper with $\varGamma_\mathrm{eff}$ and $\kappa_\mathrm{eff}$ as fit parameters.

\subsection{Equations of motion with noise}

To treat the sideband-cooling of the system, we have to include the input noise terms again and get as starting point (when the only relevant coupling is given by $g_-$)
\begin{eqnarray}
	\frac{\hat{c}_0}{\chi_{\mathrm{p}, 0}} & = & -ig_- \hat{b}_0^\dagger + i\mathcal{K}n_\mathrm{d}\hat{c}_2^\dagger + \sqrt{\kappa_\mathrm{e}}\hat{\xi}_\mathrm{e, 0} + \sqrt{\kappa_\mathrm{i}}\hat{\xi}_\mathrm{i, 0} \\
	\frac{\hat{c}_2^\dagger}{\overline{\chi}_{\mathrm{p}, 2}} & = & - i\mathcal{K}n_\mathrm{d}\hat{c}_0 + \sqrt{\kappa_\mathrm{e}}\hat{\xi}_\mathrm{e, 2}^\dagger + \sqrt{\kappa_\mathrm{i}}\hat{\xi}_\mathrm{i, 2}^\dagger \\
	\frac{\hat{b}_0^\dagger}{\overline{\chi}_{0, 0}} & = & ig_-^*\hat{c}_0 + \sqrt{\varGamma_\mathrm{e}}\hat{\zeta}_\mathrm{e, 0}^\dagger + \sqrt{\varGamma_\mathrm{i}}\hat{\zeta}_\mathrm{i, 0}^\dagger
\end{eqnarray}
with (we write it down again here for clarity)
\begin{equation}
	\chi_\mathrm{p, 0} = \frac{1}{\frac{\kappa}{2} + i\left(\varDelta_\mathrm{p} - 2\mathcal{K}n_\mathrm{d} + \varOmega \right)}, ~~~ \overline{\chi}_\mathrm{p, 2} = \frac{1}{\frac{\kappa}{2} - i\left(\varDelta_\mathrm{p} - 2\mathcal{K}n_\mathrm{d} - \varOmega + 2\varOmega_\mathrm{dp} \right)}, ~~~ \overline{\chi}_{0, 0} = \frac{1}{\frac{\varGamma_0}{2} + i\left( \varOmega + \varOmega_0 \right)}.
\end{equation}
After elimination of $\hat{c}_2^\dagger$ and the approximation of $\chi_\mathrm{g} \approx \chi_\mathcal{G}$ around the idler resonance the equations read
\begin{eqnarray}
	\frac{\hat{c}_0}{\chi_\mathcal{G}} & = & -ig_- \hat{b}_0^\dagger + \sqrt{\kappa_\mathrm{e}}\left[\hat{\xi}_\mathrm{e, 0} + i\mathcal{K}n_\mathrm{d}\overline{\chi}_\mathrm{p, 2} \hat{\xi}_\mathrm{e, 2}^\dagger \right] + \sqrt{\kappa_\mathrm{i}}\left[\hat{\xi}_\mathrm{i, 0}  + i\mathcal{K}n_\mathrm{d}\overline{\chi}_\mathrm{p, 2} \hat{\xi}_\mathrm{i, 2}^\dagger\right] \label{eqn:c0_w_noise}\\
	\frac{\hat{b}_0^\dagger}{\overline{\chi}_{0, 0}} & = & ig_-^*\hat{c}_0 + \sqrt{\varGamma_\mathrm{e}}\hat{\zeta}_\mathrm{e, 0}^\dagger + \sqrt{\varGamma_\mathrm{i}}\hat{\zeta}_\mathrm{i, 0}^\dagger. \label{eqn:b0_w_noise}
\end{eqnarray}
These are the equations of motion of a usual photon-pressure system with a scaled susceptibility and with a modified cavity bath.
Note that we keep the indices to make explicitly clear that the noise operators $\hat{\xi}_{0}$ are the noise Fourier components at a different frequency than $\hat{\xi}_2^\dagger$.

\subsection{HF cavity output field}

When we combine the equations (\ref{eqn:c0_w_noise}) and (\ref{eqn:b0_w_noise}), the equation for $\hat{c}_0$ becomes 
\begin{eqnarray}
	\frac{\hat{c}_0}{\chi_\mathcal{G}^\mathrm{eff}} & = & -ig_- \overline{\chi}_{0, 0} \sqrt{\varGamma_\mathrm{e}}\hat{\zeta}_\mathrm{e, 0}^\dagger -ig_- \overline{\chi}_{0, 0} \sqrt{\varGamma_\mathrm{i}}\hat{\zeta}_\mathrm{i, 0}^\dagger + \sqrt{\kappa_\mathrm{e}}\left[\hat{\xi}_\mathrm{e, 0} + i\mathcal{K}n_\mathrm{d}\overline{\chi}_\mathrm{p, 2} \hat{\xi}_\mathrm{e, 2}^\dagger \right] + \sqrt{\kappa_\mathrm{i}}\left[\hat{\xi}_\mathrm{i, 0}  + i\mathcal{K}n_\mathrm{d}\overline{\chi}_\mathrm{p, 2} \hat{\xi}_\mathrm{i, 2}^\dagger\right]
\end{eqnarray}
with the effective cavity susceptibility
\begin{equation}
	\chi_\mathcal{G}^\mathrm{eff} = \frac{\chi_\mathcal{G}}{1 - |g_-|^2\chi_\mathcal{G}\overline{\chi}_{0, 0}}
\end{equation}
To calculate the idler resonance output field, we use
\begin{equation}
	\hat{c}_\mathrm{out} = \hat{\xi}_\mathrm{e, 0} - \sqrt{\kappa_\mathrm{e}}\hat{c}_\mathrm{0}
\end{equation}
and obtain
\begin{eqnarray}
	\hat{c}_\mathrm{out}& = & ig_- \chi_\mathcal{G}^\mathrm{eff}\overline{\chi}_{0, 0} \sqrt{\kappa_\mathrm{e}}\left[\sqrt{\varGamma_\mathrm{e}}\hat{\zeta}_\mathrm{e, 0}^\dagger + \sqrt{\varGamma_\mathrm{i}}\hat{\zeta}_\mathrm{i, 0}^\dagger\right] \nonumber \\
	& & + \hat{\xi}_\mathrm{e, 0}\left[1 - \kappa_\mathrm{e}\chi_\mathcal{G}^\mathrm{eff}\right]- i\mathcal{K}n_\mathrm{d}\overline{\chi}_\mathrm{p, 2}\kappa_\mathrm{e}\chi_\mathcal{G}^\mathrm{eff} \hat{\xi}_\mathrm{e, 2}^\dagger - \sqrt{\kappa_\mathrm{i}\kappa_\mathrm{e}}\chi_\mathcal{G}^\mathrm{eff}\left[\hat{\xi}_\mathrm{i, 0}  + i\mathcal{K}n_\mathrm{d}\overline{\chi}_\mathrm{p, 2} \hat{\xi}_\mathrm{i, 2}^\dagger\right].
\end{eqnarray}
For the symmetric output field power spectral density we get therefore
\begin{eqnarray}
	S_\mathrm{out} & \approx & \frac{1}{2} + n_\mathrm{e}^\mathrm{HF} + \left| g_- \chi_\mathcal{G}^\mathrm{eff}\overline{\chi}_{0, 0} \sqrt{\kappa_\mathrm{e}} \right|^2 \varGamma_0 \left[n_\mathrm{th}^\mathrm{RF} + n_\mathrm{e}^\mathrm{HF} + 1\right] - \kappa_\mathrm{e}\kappa\left|\chi_\mathcal{G}^\mathrm{eff}\right|^2\left(1 - \mathcal{K}^2 n_\mathrm{d}^2|\overline{\chi}_\mathrm{p, 2}|^2 \right) \left[ n_\mathrm{e}^\mathrm{HF} + \frac{1}{2}  \right] \nonumber \\
	& & + \kappa_\mathrm{e}^2\left|\chi_\mathcal{G}^\mathrm{eff}\right|^2\left[ n_\mathrm{e}^\mathrm{HF} + \frac{1}{2}  \right] + \kappa_\mathrm{e}\kappa_\mathrm{i}\left|\chi_\mathcal{G}^\mathrm{eff}\right|^2\left[ n_\mathrm{i}^\mathrm{HF} + \frac{1}{2}  \right] + \kappa_\mathrm{e}\kappa\mathcal{K}^2n_\mathrm{d}^2|\overline{\chi}_\mathrm{p, 2}|^2\left| \chi_\mathcal{G}^\mathrm{eff} \right|^2 \left[ n_\mathrm{th}^\mathrm{HF} + \frac{1}{2} \right] \\
	& = & \frac{1}{2} + n_\mathrm{e}^\mathrm{HF} + \left| g_- \chi_\mathcal{G}^\mathrm{eff}\overline{\chi}_{0, 0} \sqrt{\kappa_\mathrm{e}} \right|^2 \varGamma_0 \left[n_\mathrm{th}^\mathrm{RF} + n_\mathrm{e}^\mathrm{HF} + 1\right] + \kappa_\mathrm{e}\kappa_\mathrm{i}\left|\chi_\mathcal{G}^\mathrm{eff}\right|^2\left[ n_\mathrm{i}^\mathrm{HF} -  n_\mathrm{e}^\mathrm{HF} \right] \nonumber \\
	& &  + \kappa_\mathrm{e}\kappa_\mathrm{i}\mathcal{K}^2n_\mathrm{d}^2|\overline{\chi}_\mathrm{p, 2}|^2\left| \chi_\mathcal{G}^\mathrm{eff} \right|^2 \left[ n_\mathrm{e}^\mathrm{HF} + n_\mathrm{i}^\mathrm{HF} +  1 \right] + \kappa_\mathrm{e}^2\mathcal{K}^2n_\mathrm{d}^2|\overline{\chi}_\mathrm{p, 2}|^2\left| \chi_\mathcal{G}^\mathrm{eff} \right|^2 \left[ 2n_\mathrm{e}^\mathrm{HF} + 1 \right]
\end{eqnarray}
where we used 
\begin{equation}
	n_\mathrm{th}^\mathrm{RF} = \frac{\varGamma_\mathrm{i}n_\mathrm{i}^\mathrm{RF} + \varGamma_\mathrm{e}n_\mathrm{e}^\mathrm{RF}}{\varGamma_0}, ~~~~~ n_\mathrm{th}^\mathrm{HF} = \frac{\kappa_\mathrm{i}n_\mathrm{i}^\mathrm{HF} + \kappa_\mathrm{e}n_\mathrm{e}^\mathrm{HF}}{\kappa_0}
\end{equation}
For vanishing thermal occupation in the HF domain $n_\mathrm{i}^\mathrm{HF} = n_\mathrm{i}^\mathrm{HF} \approx 0$ and including the added noise of the HEMT amplifier $n_\mathrm{add}$, we get
\begin{eqnarray}
	S_\mathrm{out}^\mathrm{q} & \approx & \frac{1}{2} + n_\mathrm{add} +  \kappa_\mathrm{e}\left| g_- \right|^2 \left|\chi_\mathcal{G}^\mathrm{eff} \right|^2\left|\overline{\chi}_{0, 0} \right|^2 \varGamma_0 \left(n_\mathrm{th}^\mathrm{RF} + 1\right) + \kappa_\mathrm{e}\kappa\mathcal{K}^2n_\mathrm{d}^2|\overline{\chi}_\mathrm{p, 2}|^2\left| \chi_\mathcal{G}^\mathrm{eff} \right|^2
	\label{eqn:FullNLPSD}
\end{eqnarray}

\subsection{Residual and cooled occupation}

To calculate the residual occupation of the HF and RF mode under idler-mode blue-sideband pumping, we first express the HF and RF operators as functions of the input noise terms only and get
\begin{eqnarray}
	\frac{\hat{c}_0}{\chi_\mathcal{G}^\mathrm{eff}} & = & -ig_- \overline{\chi}_{0, 0} \sqrt{\varGamma_\mathrm{e}}\hat{\zeta}_\mathrm{e, 0}^\dagger -ig_- \overline{\chi}_{0, 0} \sqrt{\varGamma_\mathrm{i}}\hat{\zeta}_\mathrm{i, 0}^\dagger + \sqrt{\kappa_\mathrm{e}}\left[\hat{\xi}_\mathrm{e, 0} + i\mathcal{K}n_\mathrm{d}\overline{\chi}_\mathrm{p, 2} \hat{\xi}_\mathrm{e, 2}^\dagger \right] + \sqrt{\kappa_\mathrm{i}}\left[\hat{\xi}_\mathrm{i, 0}  + i\mathcal{K}n_\mathrm{d}\overline{\chi}_\mathrm{p, 2} \hat{\xi}_\mathrm{i, 2}^\dagger\right] \\
	\frac{\hat{b}_0^\dagger}{\overline{\chi}_{0, 0}^\mathrm{eff}} & = & ig_-^*\chi_\mathcal{G}\sqrt{\kappa_\mathrm{e}}\left[\hat{\xi}_\mathrm{e, 0} + i\mathcal{K}n_\mathrm{d}\overline{\chi}_\mathrm{p, 2} \hat{\xi}_\mathrm{e, 2}^\dagger \right] + ig_-^*\chi_\mathcal{G}\sqrt{\kappa_\mathrm{i}}\left[\hat{\xi}_\mathrm{i, 0}  + i\mathcal{K}n_\mathrm{d}\overline{\chi}_\mathrm{p, 2} \hat{\xi}_\mathrm{i, 2}^\dagger\right]  + \sqrt{\varGamma_\mathrm{e}}\hat{\zeta}_\mathrm{e, 0}^\dagger + \sqrt{\varGamma_\mathrm{i}}\hat{\zeta}_\mathrm{i, 0}^\dagger.
\end{eqnarray}
with
\begin{equation}
	\chi_\mathcal{G}^\mathrm{eff} = \frac{\chi_\mathcal{G}}{1 - |g_-|^2\chi_\mathcal{G}\overline{\chi}_{0, 0}}, ~~~ \overline{\chi}_{0, 0}^\mathrm{eff} = \frac{\overline{\chi}_{0, 0}}{1 - |g_-|^2\chi_\mathcal{G}\overline{\chi}_{0, 0}}
\end{equation}
From here, we can calculate the power spectral densities of the two modes via $S_n^\mathrm{HF} = \langle \hat{c}_0^\dagger \hat{c}_0 \rangle$ and $S_n^\mathrm{RF} = \langle \hat{b}_0^\dagger \hat{b}_0 \rangle$, respectively.
We get for the high-frequency mode
\begin{equation}
	S_n^\mathrm{HF} = \varGamma_0|g_-|^2 |\chi_\mathcal{G}^\mathrm{eff}|^2|\overline{\chi}_{0, 0}|^2\left( n_\mathrm{th}^\mathrm{RF} + 1\right) + \kappa |\chi_\mathcal{G}^\mathrm{eff}|^2 n_\mathrm{th}^\mathrm{HF} + \kappa\mathcal{K}^2 n_\mathrm{d}^2 |\overline{\chi}_\mathrm{p, 2}|^2 |\chi_\mathcal{G}^\mathrm{eff}|^2 \left( n_\mathrm{th}^\mathrm{HF} + 1 \right)
\end{equation}
and for the RF mode
\begin{equation}
	S_n^\mathrm{RF} = \varGamma_0|\overline{\chi}_{0, 0}^\mathrm{eff}|^2  n_\mathrm{th}^\mathrm{RF} + \kappa|g_-|^2|\overline{\chi}_{0, 0}^\mathrm{eff}|^2|\chi_\mathcal{G}|^2  n_\mathrm{th}^\mathrm{HF} + \kappa|g_-|^2\mathcal{K}^2 n_\mathrm{d}^2 |\overline{\chi}_\mathrm{p, 2}|^2 |\overline{\chi}_{0, 0}^\mathrm{eff}|^2 |\chi_\mathcal{G}|^2  \left( n_\mathrm{th}^\mathrm{HF} + 1 \right).
\end{equation}
We can rewrite this as
\begin{eqnarray}
	S_n^\mathrm{HF} & = & \varGamma_0|g_-|^2 |\chi_\mathcal{G}^\mathrm{eff}|^2|\overline{\chi}_{0}|^2\left( n_\mathrm{th}^\mathrm{RF} + 1\right) + \kappa |\chi_\mathcal{G}^\mathrm{eff}|^2\left[ n_\mathrm{th}^\mathrm{HF} + \frac{\mathcal{G}-1}{\mathcal{G}} \left( n_\mathrm{th}^\mathrm{HF} + 1 \right)\right] \label{eqn:PSD_HF}\\
	S_n^\mathrm{RF} & = & \varGamma_0|\overline{\chi}_{0}^\mathrm{eff}|^2  n_\mathrm{th}^\mathrm{RF} + \kappa|g_-|^2|\overline{\chi}_{0}^\mathrm{eff}|^2|\chi_\mathcal{G}|^2  \left[n_\mathrm{th}^\mathrm{HF} + 1 + \frac{\mathcal{G}-1}{\mathcal{G}} n_\mathrm{th}^\mathrm{HF}\right],
	\label{eqn:PSD_RF}
\end{eqnarray}
where we also omitted the subscripts indicating the particular frequency range as it is not needed anymore in this formulation.
By integration, we find now for the final occupation of both modes
\begin{eqnarray}
	n_\mathrm{fin}^\mathrm{HF} & = & \frac{\kappa}{\kappa + \varGamma_0}\frac{4g_\mathrm{eff}^2 + \varGamma_0(\kappa + \varGamma_0)}{4g_\mathrm{eff}^2 + \kappa\varGamma_0}\left|\mathcal{G}\right|\left(\tilde{n}_\mathrm{eff}^\mathrm{HF} + 1\right) + \frac{\varGamma_0}{\kappa + \varGamma_0}\frac{4g_\mathrm{eff}^2}{4g_\mathrm{eff}^2 + \kappa\varGamma_0}\left|\mathcal{G}\right|\left(n_\mathrm{th}^\mathrm{RF} + 1\right) \label{eqn:nfin_HF} \\
	n_\mathrm{fin}^\mathrm{RF} & = & \frac{\varGamma_0}{\kappa + \varGamma_0}\frac{4g_\mathrm{eff}^2 + \kappa(\kappa + \varGamma_0)}{4g_\mathrm{eff}^2 + \kappa\varGamma_0}n_\mathrm{th}^\mathrm{RF} + \frac{\kappa}{\kappa + \varGamma_0}\frac{4g_\mathrm{eff}^2}{4g_\mathrm{eff}^2 + \kappa\varGamma_0}\tilde{n}_\mathrm{eff}^\mathrm{HF}
	\label{eqn:nfin_RF}
\end{eqnarray}
with
\begin{equation}
	g_\mathrm{eff}^2 = |\mathcal{G}||g_-|^2, ~~~~~ \tilde{n}_\mathrm{eff}^\mathrm{HF} = n_\mathrm{th}^\mathrm{HF}\left( 2|\mathcal{G}| + 1 \right) + |\mathcal{G}|,
\end{equation}
where $\tilde{n}_\mathrm{eff}^\mathrm{HF}$ was chosen to be by definition the effective thermal cavity occupation seen by the RF mode.
We can also calculate the limit occupation of both modes for $g_\mathrm{eff} \gg \kappa, \varGamma_0$ and for the HF cavity being in the ground-state $n_\mathrm{th}^\mathrm{HF}$ and get
\begin{eqnarray}
	n_\mathrm{lim}^\mathrm{HF} & = & \frac{\kappa}{\kappa + \varGamma_0}\left|\mathcal{G}\right|\left(\left|\mathcal{G}\right| + 1 \right) + \frac{\varGamma_0}{\kappa + \varGamma_0}\left|\mathcal{G}\right|\left(n_\mathrm{th}^\mathrm{RF}+1 \right) \\
	n_\mathrm{lim}^\mathrm{HF} & = & \frac{\varGamma_0}{\kappa + \varGamma_0}n_\mathrm{th}^\mathrm{RF} + \frac{\kappa}{\kappa + \varGamma_0}\left|\mathcal{G}\right|.
\end{eqnarray}
For our particular experimental situation presented in Fig.~3 of the main paper this means that $n_\mathrm{lim}^\mathrm{RF} \approx 2.1$ and $n_\mathrm{lim}^\mathrm{HF} \approx 1.1$.
The two values are shown there by dashed gray lines.

\section{Supplemental Note 2: Adjusting $\mathcal{G}$ with drive power}
\begin{figure}[h]
	\centerline{\includegraphics[trim = {4cm, 4.5cm, 3cm, 2cm}, clip=True,width=0.95\textwidth] {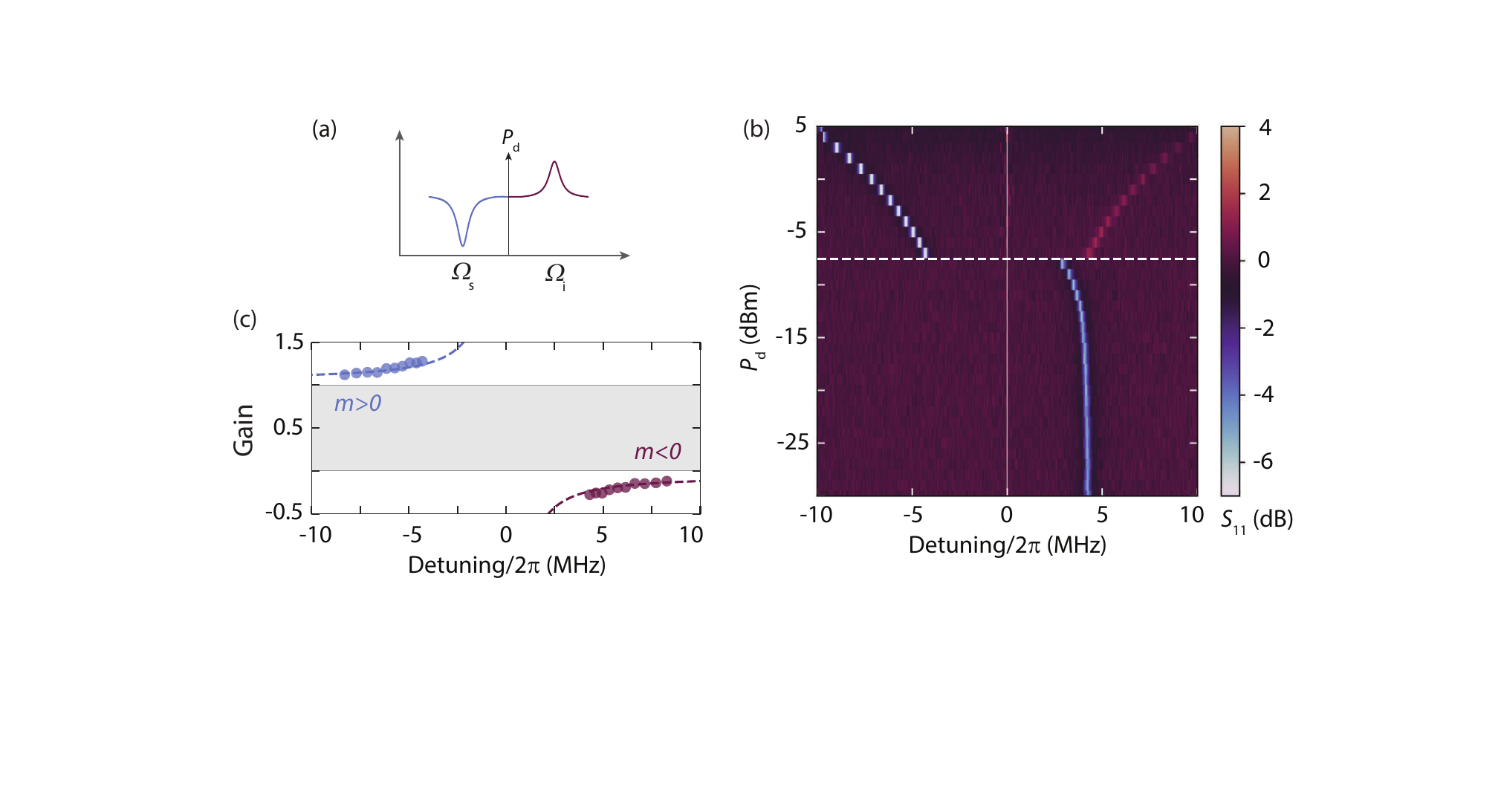}}
	\caption{\textsf{\textbf{Drive-power-tuning of the gain factor $\mathcal{G}$.} (a) shows a schematic of the experiment. A drive tone with variable power $P_\mathrm{d}$ is applied slightly red-detuned from the bare cavity. As a result the single cavity mode splits into two modes, a signal and an idler mode with resonance frequencies $\varOmega_\mathrm{i} = \varOmega_0$ and $\varOmega_\mathrm{s} = -\varOmega_0$, respectively. Resnance frequencies are given with respect to the drive tone. A small probe tone measures the reflection response for each drive power. The response for frequencies $\omega < \omega_\mathrm{d}$ is colored in light blue and corresponds to the positive mass mode, the response for $\omega > \omega_\mathrm{d}$ is colored in purple and corresponds to the negative mass mode. (b) Color-coded reflection response $S_{11}$ in dB vs probe frequency (detuning from the drive $\omega - \omega_\mathrm{d}$) and drive power $P_\mathrm{d}$. The bifurcation point is marked with a horizontal dashed line; for powers below that point there is only one mode visible, for powers above there are two modes. From fits to each linescan using Eq.~\ref{eqn:2ModeFit}, we obtain as one of the fit parameters the gain factor $\mathcal{G}$, which is plotted in panel (c) vs detuning from the drive tone. The panel shows $\mathcal{G}$ in purple for the negative mass mode (idler resonance) and $1 - \mathcal{G}$ in light blue for the positive mass mode (signal resonance). The gray area covers the range not accessible through parametric driving in either of the modes. The dashed lines are theory curves following Eq.~(\ref{eqn:G}). }}
	\label{fig:PDep}
\end{figure}
The factor $\mathcal{G}$ describing the generalized susceptibility can be adjusted in our system with drive power and drive detuning from the bare resonance frequency as described by Eq.~(\ref{eqn:G}).
To show exemplarily how $\mathcal{G}$ is modified with drive power, we recorded reflection spectra at varying drive power and a fixed detuning from the bare resonance and observe a similar behaviour as the one reported in Ref.~\cite{FaniSani21_SI}.
The results of one of these measurements is shown in Supplemental Fig.~\ref{fig:PDep}.
For a red-detuned drive and low drive powers the undriven cavity is unperturbed; then for increasing powers it starts to get shifted towards the drive and at the bifurcation power threshold, the cavity jumps to a high amplitude state, revealing two modes in the probe reflection, located symmetrically around the drive, cf. Supplemental Fig.~\ref{fig:PDep}(b).
With further increasing power the signal mode (left, dip) is shifting towards lower frequencies, while the idler mode (right, peak) is shifting equivalently towards higher frequencies.
While the two modes are shifting, also their amplitude (dip-depth, peak-height) changes, indicating a power-dependent $\mathcal{G}$.
From fits to each linescan of the reflection in Supplemental Fig.~\ref{fig:PDep}(b) above the bifurcation point and using a linear two-mode model
\begin{equation}
	S_{11} = 1 - \kappa_\mathrm{e}\left(1-\mathcal{G}\right)\chi_\mathrm{s} - \kappa_\mathrm{e}\mathcal{G}\chi_\mathrm{i}
	\label{eqn:2ModeFit}
\end{equation}
where
\begin{equation}
	\chi_\mathrm{s} = \frac{1}{\frac{\kappa}{2} + i\left( \omega - \omega_\mathrm{d} - \varOmega_0 \right)}, ~~~~~ \chi_\mathrm{i} = \frac{1}{\frac{\kappa}{2} + i\left( \omega - \omega_\mathrm{d} + \varOmega_0 \right)}
\end{equation}
we obtain as fit parameter $\kappa, \varOmega_0$ and in particular the corresponding $\mathcal{G}$, the result is plotted in (c) together with a line describing the theoretical expectation Eq.~(\ref{eqn:G}).
By just replacing $\varOmega_\mathrm{i} \rightarrow \varOmega_\mathrm{s} = -\varOmega_\mathrm{i}$ in Eq.~(\ref{eqn:G}), we can also get a theoretical line for the gain factor of the signal mode.
Both datasets, idler and signal $\mathcal{G}$, show good agreement between theory and fit values.
For the data presented in the main paper, we chose somewhat different operation points for (drive powers/detunings) than the ones shown in Supplemental Fig.~\ref{fig:PDep}.
It is useful, however, to quickly discuss some boundary conditions for choosing suitable $\mathcal{G}$s.
We wanted to work in a regime where $\mathcal{G}$ is in good approximation a real-valued number, which limits us to the regime $\varOmega_\mathrm{i} \gg \kappa$.
For the future, however, it would be also very interesting to work in a regime where $\mathcal{G}$ is complex-valued, i.e., the HF mode has a complex-valued mass.
We also want to work in a regime, where the gain does not vary too much in the relevant frequency range, so we can set it as a constant in the relevant equations.
At the same time, we want it to be as large as possible to get clear signals in the VNA and spectrum analyzer measurements.
For most main paper datsets, the relevant frequency span is on the order of two cavity linewidths, while for the normal-mode splitting in Fig.~2(d) and (e) we naturally have more relevant frequency span of $\sim 4\kappa$ due to the split modes.
Therefore we chose to measure the normal-mode splitting with smaller gain $\mathcal{G} \approx -0.2$ than the other parts.
Note in Supplemental Fig.~\ref{fig:PDep}(c) how the slope of $\mathcal{G}(\omega)$ gets smaller with decreasing $|\mathcal{G}|$.
As a consequence it would be straightforward to operate at $\mathcal{G} = -1$ if one could obtain a mode with a smaller linewidth.

\section{Supplemental Note 3: Data analysis and parameter extraction}
\subsection{General}
All data presented in the manuscript are processed using python scripts.
The shown $S_{11}$ data are almost all fitted with a generalized function
\begin{equation}
	S_{11}^\mathrm{fit}(\omega) = \left(a_0 + a_1\omega + a_2\omega^2\right)\left[1 - f(\omega)e^{i\theta}\right]e^{i\left(\phi_0 + \phi_1\omega\right)}
\end{equation}
where $\omega$ is the probe angular frequency, $a_j$ are the amplitude background coefficients, $\phi_j$ are the phase background coefficients, $f(\omega)$ is the complex-valued response function of the device and $\theta$ is a rotational fit factor taking into account Fano-like interference effects in the setup.
The only exception is the theory line in Fig.~2(e), which is not a fit at all.
\subsection{Figure 1}
For the fit lines in main paper Fig.~1(c), the response function is given by
\begin{equation}
	f(\omega) = \kappa_\mathrm{e}\chi_\mathcal{G}
\end{equation}
where for the undriven case $\mathcal{G} = 1$ and for the driven signal and idler modes it is a fit parameter.
The resulting $\chi_\mathcal{G}$ is also what is plotted as theory/fit line in (d) and (e).
In addition, but not shown, we fit the data with the full Kerr nonlinear equation
\begin{equation}
	f(\omega) = \kappa_\mathrm{e}\chi_\mathrm{g},
\end{equation}
cf. Eq.~(\ref{eqn:FullKerr}) in order to obtain values for $\mathcal{K} = -2\pi\cdot6.609\,$kHz and the intracircuit drive photon number $n_\mathrm{d} = 991$, which we will use in the analysis of the cooling data shown in Fig.~3.
\subsection{Figure 2}
For the fits in panel (b) of main paper Fig.~2, displaying photon-pressure induced absorption (equivalent to electromagnetically induced absorption and optomechanically induced absorption), we use as obtained from Eq.~(\ref{eqn:Fig2bfunc})
\begin{equation}
	f(\omega) = \mathcal{G}\frac{\frac{\varGamma_0}{2} + i\left( \varOmega + \varOmega_0 \right)}{\left[\frac{\varGamma_\mathrm{eff}}{2} + i\left(\varOmega + \varOmega_0 \right) \right] \left[ \frac{\kappa_\mathrm{eff}}{2} + i\left(\varOmega + \varOmega_0 \right)  \right]}
\end{equation}
where $\varGamma_0$ and $\varOmega_0$ are fixed parameters, $\varOmega$ is the scan frequency relative to the sideband-pump and $\varGamma_\mathrm{eff}$ and $\kappa_\mathrm{eff}$ are the main two fit parameters and we allow for small variations of $\mathcal{G}$ between $-0.33$ and $-0.36$, since the sideband pump tone seems to slightly change the gain.
In addition, we set as boundary condition $\kappa_\mathrm{eff} < \kappa$ and $\varGamma_\mathrm{eff}> \varGamma_0$.
The two fit parameters $\varGamma_\mathrm{eff}$ and $\kappa_\mathrm{eff}$ as obtained from the fit function are then plotted in panel (c).
We note that for each power we have measured the same $S_{11}$ three times and we therefore perform three individual fits for each pump power, the values for $\varGamma_\mathrm{eff}$ and $\kappa_\mathrm{eff}$ in panel (c) are the mean of the three individual fits.
In order to demonstrate how $\varGamma_\mathrm{eff}$ and $\kappa_\mathrm{eff}$ manifest in the reflection $S_{11}$, we plot two sets of theoretical curves in Supplemental Fig.~\ref{fig:FigureS2}, one for constant $\kappa_\mathrm{eff}$ and varying $\varGamma_\mathrm{eff}$ and one vice versa based on Eq.~(\ref{eqn:Fig2bfunc}).
\begin{figure}[h]
	\centerline{\includegraphics[trim = {4.5cm, 6cm, 4cm, 2.8cm}, clip=True,width=0.85\textwidth] {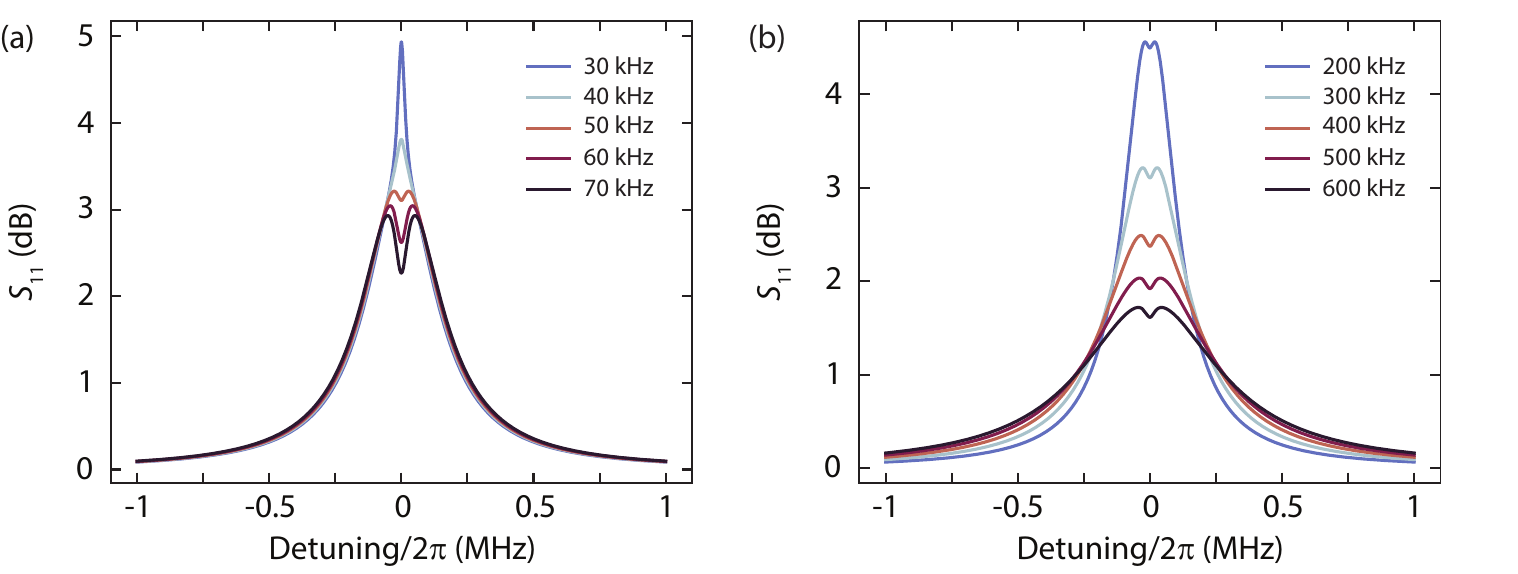}}
	\caption{\textsf{\textbf{How $\varGamma_\mathrm{eff}$ and $\kappa_\mathrm{eff}$ manifest in the reflection $S_{11}$ in the weak-coupling regime.} Theoretical curves for the reflection $S_{11}$ for (a) constant $\kappa_\mathrm{eff} = 2\pi\cdot 300\,$kHz and varying $\varGamma_\mathrm{eff} = 30...70\,$kHz (color-code explained by legend) and for (b) constant $\varGamma_\mathrm{eff} = 2\pi\cdot 50\,$kHz and varying $\kappa_\mathrm{eff} = 200...600\,$kHz (color-code explained by legend). The other parameters are $\mathcal{G} = -0.35$, $\varGamma_0 = 2\pi\cdot45\,$kHz, and $\varOmega_0 = 2\pi\cdot450\,$MHz. For constant $\kappa_\mathrm{eff}$ (a) the broad resonance stays nearly unmodified with varying $\varGamma_\mathrm{eff}$, the most prominent signature of changing $\varGamma_\mathrm{eff}$ is a change of the depth of the narrow dip inside the broad one. Note that for $\varGamma_\mathrm{eff} < \varGamma_0$ the dip turns into a peak and that for $\varGamma_\mathrm{eff} = \varGamma_0$ the signature of the narrow Lorentzian would be completely gone as is also visible from Eq.~(\ref{eqn:Fig2bfunc}), since the terms containing $\varGamma_\mathrm{eff}$ and $\varGamma_0$ in the numerator and denominator, respectively, would just cancel. For constant $\varGamma_\mathrm{eff}$ and varying $\kappa_\mathrm{eff}$ as discussed in (b), the small dip at the peak of the broad Lorentzian is remaining nearly unmodified, while the width and height of the broad Lorentzian are clearly following the variation of $\kappa_\mathrm{eff}$.}}
	\label{fig:FigureS2}
\end{figure}
From these plots it becomes clear that in the regime where $\varGamma_\mathrm{eff} \ll \kappa_\mathrm{eff}$ the two parameters really have significantly different impacts on the reflection parameter and can be reliably extracted, as is also suggested by the good agreement between the fit parameters and the analytical theoretical lines in main paper Fig.~2(c).
Once the system comes close to the normal-mode-splitting point, however, and $\varGamma_\mathrm{eff} \sim \kappa_\mathrm{eff}$, the overall response is only very slightly depending on the exact values and it becomes much harder to extract reliable numbers with the method we present here.
The line shown in panel (e), the normal-mode-splitting regime, is a theoretical line (not a fit), where we used 
\begin{equation}
	f(\omega) = \frac{\kappa_\mathrm{e}\chi_\mathcal{G}}{1 - |g_-|^2\chi_\mathcal{G}\overline{\chi}_0}
\end{equation}
with the parameters $\omega_0 = 2\pi\cdot 7.2134\,$GHz, $\kappa = 2\pi\cdot290\,$kHz, $\kappa_\mathrm{e} = 2\pi\cdot 78\,$kHz, $\varGamma_0 = 2\pi\cdot45\,$kHz, $\varOmega_0 = 2\pi\cdot452.6\,$MHz, $G = -0.21$ and $g_- = 2\pi\cdot 480.05\,$kHz.
Most of these numbers we obtain from the fits in Fig.~1, and $g_- = \sqrt{n_-}g_0$ is calculated via the theoretical $g_0$ and the pump photon number $n_-$ calculated through the microwave generator output power and the attenuation of the input line.
\subsection{Figure 3}
For the analysis of the thermal noise data, we start by extracting a large set of parameters from the dataset discussed in Fig.~2, but up to larger powers and by fitting the data with the usual equation
\begin{equation}
	f(\omega) = \frac{\kappa_\mathrm{e}\chi_\mathcal{G}}{1 - |g_-|^2\chi_\mathcal{G}\overline{\chi}_0}.
\end{equation}
Also, we do not do complex fitting here and only work with a constant background, i.e.,
\begin{equation}
	|S_{11}| = a\big|1 - f(\omega) \big|
\end{equation}
with a constant, real-valued $a > 0$.
Again, we fit three subsequently recorded $S_{11}$ datasets for each pump power and give the mean of the fit parameters to the fit function for the power spectral density.
As totally fixed values we use $g_0$, $n_-$, $\mathcal{G}$, $\kappa_\mathrm{e}$ and $\varOmega_0$.
As starting parameters for the fit we use the values obtained from the Fig.~1 fits and allow for small variations of the HF mode resonance frequency $\omega_0 \pm 2\pi\cdot 100\,$kHz, the total linewidth $\kappa \pm 2\pi\cdot 20\,$kHz, and $\varGamma_0 - 2\pi\cdot2\,$kHz. 
As unconstrained fit parameter we have the background offset $a$.
Limiting the range of the fit parameters turned out to be important to have a reliable convergence of the fit.
As next step, we inject all parameters into the full thermal noise equation Eq.~(\ref{eqn:FullNLPSD}), the only remaining fit parameters are $n_\mathrm{th}^\mathrm{RF}$ and the white background noise floor.
Note, that using the full equation with frequency-dependent gain instead of an approximated version with constant $\mathcal{G}$ was required to account for the asymmetries observable in the noise spectra for the highest powers.
We then use the relevant quantities and inject them fit by fit into Eqs.~(\ref{eqn:PSD_HF}, \ref{eqn:PSD_RF}) and numerically integrate over the resulting individual mode power spectral densities to obtain the cooled occupations.
For the gray theory lines shown in Fig.~3(c), we take the average values of all $\kappa$ and $\varGamma_0$ that we obtained from fits to the $S_{11}$ transparency data as well as a $\mathcal{G} = -0.35$ and $n_\mathrm{th}^\mathrm{RF} = 13.5$, and plot Eqs.~(\ref{eqn:nfin_RF}, \ref{eqn:nfin_HF}) for the final occupation.
The width of the theory line in the figure is given by the maximum range of $n_\mathrm{fin}^\mathrm{RF/HF}$ that we obtain when taking the standard deviations of $\kappa$ and $\varGamma_0$ as input errors for the cooled mode occupations.
So in fact it is rather a narrow area than a strict line.

\section{Supplemental Note 4: Error bars}
\subsection{Figure 2}
The error bars in main paper Fig.~2(b) originate from fitting three individual datsets for each pump power, that we took subsequently during the measurements.
The mean of the three values for $\kappa_\mathrm{eff}$ and $\varGamma_\mathrm{eff}$ is plotted as symbol and the standard deviation as the error bars.
\subsection{Figure 3}
From all the fits to the power spectral densities for all powers, we obtain a mean value for the thermal RF occupation $n_\mathrm{th}^\mathrm{RF} = 13.5$ and a standard deviation of $\sim1.1$.
In addition we obtain a standard error from each fit itself.
Subsequently, we calculate the maximum and minimum possible values for $n_\mathrm{fin}^\mathrm{HF}$ and $n_\mathrm{fin}^\mathrm{RF}$ resulting from these two errors in $n_\mathrm{th}^\mathrm{RF}$.
These errors are plotted as error bars in Fig.~3(c).
\let\oldaddcontentsline\addcontentsline
\renewcommand{\addcontentsline}[3]{}


\begin{thebibliography}{45}
\bibitem[]{Footnote1}
	\bibinfo{title}{These relations are formally identical to a standard optomechanical or photon-pressure system with a blue sideband pump. However, there are two differences to the usual case. First, we have a generalized susceptibility $\chi_\mathcal{G}$ in the equation for $\hat{c}$ with $\mathcal{G} < 0$, and secondly we have effective input noise operators $\hat{\xi}_\mathrm{e, eff}, \hat{\xi}_\mathrm{i, eff}$ with $\left\langle \hat{\xi}_\mathrm{eff}^\dagger\hat{\xi}_\mathrm{eff} \right\rangle = n_\mathrm{th}^\mathrm{HF}+\frac{\mathcal{G}-1}{ \mathcal{G}}\left(n_\mathrm{th}^\mathrm{HF} + 1\right)$ and $\left\langle \hat{\xi}_\mathrm{eff}\hat{\xi}_\mathrm{eff}^\dagger \right\rangle = n_\mathrm{th}^\mathrm{HF} + 1 + \frac{\mathcal{G}-1}{ \mathcal{G}}n_\mathrm{th}^\mathrm{HF}$ for the HF cavity with the undriven thermal bath occupation $n_\mathrm{th}^\mathrm{HF}$. This also implies that the strong drive generates a finite effective bath temperature even for $n_\mathrm{th}^\mathrm{HF} = 0$, i.e., from the cavity quantum bath.}
\end{thebibliography}
\end{document}